\documentclass[5p,times,twocolumn]{elsarticle}

\usepackage{amssymb}
\usepackage{amsmath}
\usepackage{graphicx}
\usepackage{booktabs}
\usepackage{makecell}
\usepackage{algorithmic}
\usepackage{algorithm}
\usepackage{url}
\usepackage{float}
\usepackage{microtype}
\usepackage[hidelinks]{hyperref}
\usepackage{orcidlink}
\usepackage[dvipsnames]{xcolor}

\makeatletter
\let\ps@pprintTitle\ps@plain
\makeatother
\hypersetup{pdftitle={Controlled periodic synchronization for efficient data-parallel training},pdfauthor={Imane Ettifouri, Mostapha Zbakh, Claude Tadonki}}
\begin{document}

\begin{frontmatter}

\title{Controlled periodic synchronization for efficient data-parallel training\tnoteref{publication}}

\tnotetext[publication]{\small Published in \textit{Future Generation Computer Systems}, 189 (2027), 108946.\newline
DOI: \url{https://doi.org/10.1016/j.future.2026.108946}\newline
\textcopyright{} 2026. This manuscript version is made available under the CC BY-NC-ND 4.0 license: \url{https://creativecommons.org/licenses/by-nc-nd/4.0/}.}

\author[1,2]{Imane Ettifouri\,\orcidlink{0009-0005-3042-911X}\corref{cor1}}
\ead{imane_ettifouri@um5.ac.ma}

\author[1]{Mostapha Zbakh\,\orcidlink{0000-0002-1408-3850}}

\author[2]{Claude Tadonki\,\orcidlink{0000-0003-1194-6400}}

\cortext[cor1]{Corresponding author at: ENSIAS, Mohammed V University in Rabat, Morocco.}

\affiliation[1]{organization={ENSIAS, Mohammed V University in Rabat}, country={Morocco}}

\affiliation[2]{organization={Centre de Recherche en Informatique (CRI), MINES Paris--PSL University},
               city={Fontainebleau},
             country={France}}

\begin{abstract}
Data-parallel training relies on frequent gradient synchronization across workers. Standard {DistributedDataParallel (DDP)} synchronizes gradients at every iteration, which is effective on fast local-area networks but increasingly sensitive to communication latency and network variability in geographically distributed environments. Periodic methods such as LocalSGD reduce synchronization frequency but rely mainly on parameter averaging, which may be insufficient when worker trajectories diverge. This paper studies synchronization frequency as a systems parameter for communication-constrained distributed training. We evaluate Controlled Periodic Data Parallelism (CPDP), a PyTorch-DDP-compatible strategy that alternates local updates with a reconciliation step combining gradient AllReduce and SlowMo parameter averaging. Experiments are conducted on Grid'5000 across intra-site clusters and a cross-site {wide-area network (WAN)} deployment spanning Nancy and Sophia with 16.6\,ms {round-trip time (RTT)}. In the main cross-site setting with a fixed base learning rate {$\eta_0=0.1$} on ResNet-50/CIFAR-100, CPDP achieves the highest peak test accuracy among the evaluated configurations. At $K=2$, CPDP improves over DDP by +2.28 percentage points but incurs additional wall-clock time. At $K=4$ in this WAN setting, CPDP improves over DDP by +2.44 percentage points while reducing average wall-clock time by 13.8\%. Direct profiling shows that exposed synchronization time at $K=4$ is roughly half that of DDP, {supporting the timing component of the observed WAN accuracy--time trade-off. This profiling did not allow us to clearly determine the cause of the difference in accuracy}. Additional experiments on ViT-S/CIFAR-100 and ResNet-50/TinyImageNet show that CPDP remains competitive with DDP and generally improves over LocalSGD. Learning-rate and SlowMo sensitivity analyses further show that large-scale periodic synchronization is sensitive to the interaction between learning rate, synchronization period, and stabilization. Overall, the results show that synchronization frequency is a practical control parameter for distributed training under communication constraints.
\end{abstract}


\end{frontmatter}

\section{Introduction}

Distributed training has become essential for scaling modern deep neural networks. 
Frameworks such as PyTorch DistributedDataParallel~\cite{li_pytorch_2020} allow models to be trained across multiple GPUs and nodes by synchronizing gradients through collective communication operations. 
While accelerator performance has improved rapidly, the scalability of distributed training is increasingly constrained by communication overhead and synchronization rigidity rather than raw computation. 
In multi-node environments, frequent gradient aggregation can dominate iteration time, reducing effective hardware utilization and limiting achievable speedup.

Recent research has explored system-level optimizations of distributed 
deep learning infrastructures, particularly in heterogeneous computing 
environments~\cite{ettifouri_hpc_ai_2024}. Liu et al.~\cite{liu_heterps_2023} proposed HeterPS, a distributed training framework that leverages reinforcement learning to schedule neural network layers across heterogeneous resources in order to improve training throughput and resource utilization. Similarly, Wang et al.~\cite{wang_gpars_2024} 
introduced GPARS, a graph-based predictive scheduling method designed to improve 
resource allocation efficiency in heterogeneous GPU clusters. These works focus primarily on resource scheduling and cluster utilization in distributed training systems.
In contrast, this work focuses on the synchronization behavior of data-parallel training itself, analyzing how synchronization frequency 
interacts with communication latency and affects convergence in distributed environments.
{Recent efficiency-oriented machine-learning studies have addressed model design, optimization behavior, and computational constraints across diverse settings~\cite{liu_dtae_cgan_2024,hou_powermobilenet_2025,liu_sailboat_2023,ettifouri_gradient_descent_2026}. Complementing these application-level efforts, this work focuses on synchronization frequency, collective communication, and parameter reconciliation as system-level efficiency mechanisms for data-parallel distributed training.}

Synchronous data-parallel training remains the default strategy for scaling deep neural networks. 
In this paradigm, each worker computes gradients locally and participates in collective communication operations; typically \textit{AllReduce} to maintain model consistency. 
While this approach preserves stable convergence behavior and simplifies reasoning about optimization dynamics, it introduces strict per-step synchronization barriers. 
Empirical analyses of distributed training workloads show that, beyond modest cluster sizes, collective communication frequently lies on the critical path, particularly in bandwidth-constrained multi-node deployments~\cite{yu_communication_2023,wei_communication_2025,li_understanding_2024,ye_deep_2024}. 
As a consequence, scaling efficiency may degrade even when local computation is well optimized.
In practice, modern frameworks such as PyTorch overlap the NCCL-based gradient AllReduce with the backward pass, so communication does not appear as a separate timing term in simple instrumentation~\cite{li_pytorch_2020}. 
This makes it difficult to isolate the communication pressure inside the backward phase without dedicated profiling. 
In cross-site WAN deployments, where bandwidth is limited and latency is high, {DDP's overlapped backward phase can exhibit substantial time inflation}, while the per-step global barrier 
exposes the optimization process to straggler effects and network jitter.

Recent systems research has focused on reducing exposed communication latency through improved scheduling and runtime-level optimizations. 
Techniques such as tensor-level reordering, GPU-driven collectives, and communication--computation overlap mechanisms~\cite{hwang_ark_2023,chen_centauri_2024,cao_crux_2024} demonstrate that careful orchestration and scheduling of collective operations can substantially improve utilization. 
However, these approaches preserve the fundamental iteration-synchronous execution model of data-parallel training. 
Gradient synchronization remains enforced at every mini-batch step, and the global optimization trajectory remains tightly coupled across workers.

These observations raise an important research question: is per-step global synchronization always necessary in practical deployments, or can synchronization frequency be treated as an execution parameter that adapts 
to communication constraints?

A related line of research investigates communication-efficient variants of distributed SGD, including local updates and periodic 
averaging~\cite{stich_local_2019,lin_dont_2020}. 
These approaches establish that reduced synchronization can preserve 
convergence under appropriate conditions. 
However, the behavior of these methods is often analyzed from an optimization-theoretic perspective, and fewer studies quantify their systems-level impact under heterogeneous or cross-site deployments within standard DDP workflows. 
Moreover, periodic averaging can introduce parameter drift between workers, particularly under aggressive synchronization delays or large learning rates.

In this work, we investigate synchronization frequency as a systems parameter in distributed training. 
We first {quantify a backward-time inflation proxy for the communication pressure inside} standard DDP execution across scales and network conditions.
We then study whether controlling synchronization frequency can improve training robustness in communication-constrained environments.

We implement and evaluate \emph{Controlled Periodic Data Parallelism}, an execution-level synchronization strategy within standard PyTorch DistributedDataParallel. 
Workers perform multiple local updates before synchronization. 
At synchronization boundaries, gradients are aggregated via AllReduce to {provide a synchronized gradient signal}, followed by parameter averaging stabilized with SlowMo momentum~\cite{wang_slowmo_2019}.
The implementation operates entirely within the DDP abstraction without modifying model architectures, optimizers, or communication libraries.

We evaluate CPDP on the Grid'5000 testbed~\cite{grid5000} under two complementary settings:
(i) intra-site strong scaling on ResNet-50/CIFAR-100 from 4 to 16 GPUs, and (ii) cross-site WAN training across two geographically separated Grid'5000 sites (Nancy and Sophia).
We further validate generality on TinyImageNet and Vision Transformer (ViT-S) architectures.
In all configurations, CPDP is compared against both standard DDP and LocalSGD{, with all methods using identical hyperparameters within each configuration. Learning-rate scaling follows the protocol detailed in  Section~\ref{sec:experimental_setup}.}

We do not claim that CPDP consistently outperforms fully synchronous DDP in homogeneous high-bandwidth environments.
Instead, our objective is to quantify how synchronization frequency influences system efficiency and convergence behavior, and to show that CPDP generally improves peak test accuracy over LocalSGD in the evaluated practical synchronization range while achieving higher peak test accuracy than DDP, under the fixed $\eta_0=0.1$ protocol, in the communication-constrained cross-site deployment.

Our contributions are summarized as follows:

\begin{itemize}
\item We study synchronization frequency as a controllable systems parameter in data-parallel training, focusing on communication-constrained and cross-site WAN deployments.
\item We implement CPDP as a lightweight extension of PyTorch DDP that combines periodic gradient aggregation with SlowMo-  stabilized parameter reconciliation, without modifying models, optimizers, or communication libraries.
\item We provide both an indirect DDP backward-time inflation proxy and direct exposed-synchronization profiling to characterize communication pressure across intra-site and cross-site settings.
\item We show that, under the fixed $\eta_0=0.1$ WAN protocol,
CPDP provides a favorable accuracy--time operating point, especially
at $K=4$, where it achieves higher accuracy than both baselines while
reducing wall-clock time relative to DDP.
\item We validate the behavior across additional workloads and analyze the sensitivity of periodic synchronization to $K$, learning rate, and SlowMo coefficient.
\end{itemize}

The remainder of this paper is organized as follows. 
Section~\ref{sec:related_work} reviews prior work on distributed training and communication optimization. 
Section~\ref{sec:cpdp} presents the proposed method. 
Section~\ref{sec:implementation} describes implementation details. 
Section~\ref{sec:experimental_setup} outlines the experimental methodology. 
Section~\ref{sec:results} reports results and analysis. 
Section~\ref{sec:conclusion} concludes and discusses future directions.

\section{Related Work}\label{sec:related_work}

Distributed deep neural network (DNN) training has been widely studied due to the communication overhead introduced by synchronous data-parallel execution. In this paradigm, workers repeatedly exchange gradients using collective communication operations such as \textit{AllReduce}, which can become a major scalability bottleneck as cluster size increases.
Existing efforts primarily optimize collective communication primitives, scheduling strategies, or runtime systems, while preserving the iteration-synchronous execution model. 
While these directions substantially improve communication efficiency, fewer studies explicitly examine synchronization frequency as a controllable execution parameter within standard distributed training frameworks, particularly under cross-site or WAN-constrained conditions. 
We review the most relevant directions below and position our work relative to them.

\subsection{Communication Bottlenecks in Data-Parallel Training}

Synchronous data-parallel training requires frequent gradient aggregation across workers, typically implemented using \textit{AllReduce}. 
As model size and cluster scale increase, communication overhead becomes a primary scalability constraint. 
Li et al.~\cite{li_pytorch_2020} provide a detailed characterization of PyTorch DistributedDataParallel, highlighting the tight coupling between gradient production and synchronization. 
Romero et al.~\cite{romero_accelerating_2022} demonstrate that collective communication performance often dominates the end-to-end iteration time.
More recent surveys~\cite{yu_communication_2023,wei_communication_2025} systematically analyze communication optimization techniques and conclude that communication increasingly appears on the critical path as accelerator throughput continues to increase.

Prior studies establish communication as a structural bottleneck in distributed training. However, they preserve the conventional iteration-level synchronization model: gradients are aggregated at every mini-batch step. 
As a result, while communication primitives may be optimized, the execution structure itself remains unchanged, and synchronization boundaries are not reconsidered.

\subsection{Communication Scheduling and Overlap Techniques}

A large body of systems research focuses on reducing exposed communication latency by overlapping gradient synchronization with backpropagation. 
Rashidi et al.~\cite{rashidi_enabling_2021} leverage hardware-assisted collective execution to mitigate contention. 
DeAR~\cite{zhang_dear_2023} and Mercury~\cite{duan_mercury_2022} introduce fine-grained scheduling strategies to increase overlap efficiency. 
US-Byte~\cite{gao_us-byte_2024}, PipeDAP~\cite{gao_pipedap_2025}, and D-Credit~\cite{gao_dynamic_2025} further formalize tensor-level partitioning and communication ordering to optimize overlap mathematically.

These techniques improve how communication is scheduled within each iteration. 
However, they do not reconsider whether synchronization must occur at every iteration. 
The global barrier at each mini-batch remains intact, and the optimization trajectory remains fully coupled across workers. 
Consequently, synchronization frequency itself is not treated as a tunable execution parameter.

\subsection{Advanced Communication Frameworks and Runtimes}

Several recent works focus on improving the efficiency and adaptability of collective communication libraries. 
Concerto~\cite{cheng_concerto_2025} and AutoCCL~\cite{xu_autoccl_2025} dynamically select communication strategies, while ResCCL \cite{liu_resccl_2025} improves GPU resource utilization inside collective implementations. 
Unified communication frameworks such as UCC~\cite{venkata_unified_2024} further improve portability across heterogeneous environments.

These frameworks optimize the performance of collective operations but leave the training loop structure unchanged. 
Synchronization is still enforced at each iteration, and global coordination remains mandatory before progressing to the next update. 
Thus, although collective efficiency improves, iteration-level synchronization semantics are preserved.

\subsection{{{LocalSGD}}
and Periodic Averaging Methods}\label{subsec:local_sgd}

A closely related line of research investigates periodic averaging 
strategies such as LocalSGD. 
Stich~\cite{stich_local_2019} provides convergence guarantees under 
bounded delay, showing that multiple local updates can reduce 
communication without degrading asymptotic performance. 
Lin et al.~\cite{lin_dont_2020} demonstrate that periodic 
averaging can outperform large-batch synchronous training at scale. 
Federated averaging (FedAvg)~\cite{mcmahan_communication-efficient_2023} 
further popularizes delayed synchronization in decentralized settings.

Wang et al.~\cite{wang_slowmo_2019} address the parameter 
drift problem inherent in periodic averaging by introducing SlowMo, 
a slow momentum framework applied at synchronization boundaries. 
SlowMo applies a momentum-filtered correction to the averaged 
parameters, acting as a low-pass filter over inter-worker divergence. 
The method demonstrates consistent improvements over both LocalSGD 
and decentralized SGD across image classification and machine 
translation tasks.

A separate line of work addresses distributed training across 
geographically separated sites. 
Gaia~\cite{hsieh_gaia_2017} decouples intra-datacenter communication 
from inter-datacenter communication using approximate synchronization 
to approach LAN-speed training over WAN links.   
DiLoCo \cite{douillard_diloco_2023} extends periodic synchronization 
to language model training with inner and outer optimization loops. 
These works demonstrate that cross-site training requires fundamentally 
different synchronization strategies than intra-site deployment, a 
finding consistent with our experimental observations.

While conceptually related, these approaches are primarily analyzed from an optimization-theoretic perspective. They often assume modified training semantics, decentralized control, or heterogeneous participation. Moreover, these works typically do not focus on practical integration into mainstream PyTorch DistributedDataParallel workflows.

CPDP is closely related to LocalSGD and directly builds on the 
SlowMo mechanism. Its distinguishing characteristics are operational 
rather than algorithmic. First, at synchronization boundaries, CPDP 
performs both gradient AllReduce {(providing a shared gradient input)} and parameter averaging with SlowMo correction, whereas classical LocalSGD typically performs only parameter averaging at synchronization steps. This dual-phase reconciliation combines a shared gradient signal with SlowMo-stabilized parameter averaging, at the cost of additional communication per synchronization step. Both CPDP and LocalSGD restore equality of model parameters at averaging boundaries; this property does not establish stronger consistency for CPDP nor equivalence with the DDP trajectory. Second, CPDP operates entirely within the PyTorch DDP abstraction using the \texttt{no\_sync()} context manager, without requiring custom communication libraries or modified optimizer definitions. Third, our evaluation focuses on systems-level behavior under real cross-site WAN conditions on the Grid'5000 testbed, rather than asymptotic convergence analysis.

{Importantly, after each synchronization step CPDP restores
model-parameter consistency across workers: all replicas hold identical
parameters once reconciliation completes. This is consistency at the
parameter level and does not imply equivalence to the DDP optimizer
trajectory, since optimizer internal states evolve locally between
synchronization boundaries (Section~\ref{sec:implementation}).}

\noindent\textbf{About worker scheduling and resource availability.}

Worker scheduling offers a complementary way to improve parallel execution. Li et al.~\cite{li_fault_tolerant_2023} study fault-tolerant parallel computation through network coding and scheduling of redundant tasks, allowing computation results to be recovered despite worker failures. Zheng et al.~\cite{zheng_bid_cloud_2015} study bidding for interruptible cloud resources, balancing monetary cost and completion-time requirements under spot-instance interruptions.
Leite et al.~\cite{leite2017dohko} study an autonomic approach for configuring and executing workloads across multiple clouds and provide an operational scheduler that can be considered for the distributed execution of iterative applications.
These works address which resources execute which workloads and how execution tolerates resource unavailability, whereas CPDP controls when a fixed group of training workers exchanges gradients and reconciles parameters.

The connection is that worker availability and task placement affect the cost of waiting at a synchronization boundary. Scheduling, redundancy, and resource provisioning could therefore complement synchronization-period selection. However, our experiments exclusively use reserved nodes and a fixed group of workers; CPDP does not implement coded recovery, elastic membership, or spot-instance bidding. Joint scheduling and synchronization control would require additional mechanisms for failure recovery and consistent training state, and its benefit remains to be evaluated.

\begin{table*}[htbp]
\centering
\caption{Taxonomy of representative approaches addressing communication 
overhead in distributed DNN training, categorized by optimization layer, 
execution model, and drift control mechanism.}
\label{tab:related_work_comparison}
\small
\setlength{\tabcolsep}{3pt}
\resizebox{\textwidth}{!}{%
\begin{tabular}{p{0.38\textwidth}ccccc}
\toprule

{Work category} 
& {{Communication} {scheduling}} 
& {{Library/} {Runtime}} 
& {{Execution} {reorganization}} 
& {{Iteration-} {synchronous}}
& {{Drift} {control}} \\
\midrule
DDP optimizations \cite{li_pytorch_2020,romero_accelerating_2022} 
& \checkmark & \checkmark & -- & \checkmark & -- \\

Overlap scheduling \cite{rashidi_enabling_2021,zhang_dear_2023,duan_mercury_2022} 
& \checkmark & -- & -- & \checkmark & -- \\

Tensor-level scheduling \cite{gao_us-byte_2024,gao_pipedap_2025,gao_dynamic_2025} 
& \checkmark & -- & -- & \checkmark & -- \\

Communication frameworks \cite{cheng_concerto_2025,xu_autoccl_2025,liu_resccl_2025} 
& \checkmark & \checkmark & -- & \checkmark & -- \\

LocalSGD/Periodic Avg. \cite{stich_local_2019,lin_dont_2020,mcmahan_communication-efficient_2023}
& -- & -- & \checkmark & -- & -- \\

Geo-distributed training \cite{hsieh_gaia_2017,douillard_diloco_2023}
& -- & -- & \checkmark & -- & \checkmark \\

SlowMo \cite{wang_slowmo_2019}
& -- & -- & \checkmark & -- & \checkmark \\

\textbf{CPDP (This Work, periodic synchronization study)} 
& \checkmark & -- & \checkmark & -- & \checkmark 
\\
\bottomrule
\end{tabular}%
}
\end{table*}
\subsection{Summary and Positioning}

Prior work mainly improves the efficiency of collective communication, scheduling, or runtime execution while preserving per-iteration synchronization. LocalSGD, SlowMo, and geo-distributed methods relax this constraint by reducing synchronization frequency or stabilizing periodic averaging, but they are usually studied outside standard DDP workflows or under different system assumptions.

CPDP occupies a practical middle ground. It treats synchronization frequency as an execution parameter inside PyTorch DDP, while combining gradient aggregation with SlowMo-stabilized parameter reconciliation. Table~\ref{tab:related_work_comparison} summarizes this positioning.

\section{Proposed Method}\label{sec:cpdp}

\subsection{Problem Setting}\label{subsec:problem_setting}

We consider supervised learning over dataset $\mathcal{D}$ with loss function $\ell(\theta;\xi)$ for mini-batch $\xi$ and model parameters $\theta \in \mathbb{R}^d$. Training is performed across $N$ workers. Let $\theta_{i,t}$ denote the model parameters stored on worker $i$ at training step $t$.
At training step $t$, worker $i$ samples a mini-batch
$\mathcal{B}_{i,t} \subset \mathcal{D}$ and computes the stochastic gradient
\begin{equation}
g_{i,t} \triangleq \nabla_\theta \ell(\theta_{i,t}; \mathcal{B}_{i,t}).
\end{equation}

We use $\eta_t$ for the learning rate applied at step $t$ and $\eta_0$ for the initial (base) learning rate before the warmup and cosine-annealing schedule are applied. Thus, a fixed $\eta_0$ across methods does not mean a constant learning rate throughout training.  Equations~\eqref{eq:ddp_gradient_average} and~\eqref{eq:ddp_plain_update} correspond to the plain-SGD case, omitting momentum and weight decay, and are used solely to illustrate the synchronization pattern. The experiments instead use momentum SGD for ResNet-50 and AdamW for ViT-S (Section~\ref{sec:experimental_setup}). In conventional synchronous data-parallel SGD, gradients are aggregated at every step using \texttt{AllReduce},
\begin{equation}
\bar{g}_t = \frac{1}{N} \sum_{i=1}^{N} g_{i,t},
\label{eq:ddp_gradient_average}
\end{equation}
followed by identical updates across workers:
\begin{equation}
\theta_{i,t+1} = \theta_{i,t} - \eta_t \bar{g}_t.
\label{eq:ddp_plain_update}
\end{equation}

This enforces per-step synchronization, introducing a global barrier that may dominate iteration time when communication latency is significant.

\paragraph{Boundary Conditions}

CPDP builds on optimization mechanisms previously studied in
LocalSGD~\cite{stich_local_2019} and SlowMo~\cite{wang_slowmo_2019}.
As in these works, the loss function is assumed to be smooth and the
stochastic gradients are assumed to have bounded variance.

The synchronization period satisfies $K \geq 1$; $K=1$ denotes
gradient communication at every step. Standard DDP is the baseline with
gradient synchronization at every step and without CPDP's parameter
reconciliation or SlowMo correction. Setting $K=1$ in the general CPDP
formulation alone does not recover DDP when a nonzero SlowMo correction
is retained. In the implemented warmup phase, however, $K=1$ is used
without periodic parameter reconciliation or SlowMo correction, yielding
fully synchronous per-step gradient updates.

In practice, the synchronization period and learning rate are coupled.
Larger synchronization periods increase the divergence between worker
trajectories during local-update phases, making periodic methods more
sensitive to aggressive learning-rate scaling. This behavior is observed
in the 16-GPU experiments presented in
Section~\ref{subsec:lr_sensitivity}.

If the total number of mini-batches is not divisible by $K$, the last
mini-batch of the epoch is treated as a synchronization boundary. This
ensures that all workers start the next epoch with identical parameters.

\subsection{Controlled Periodic Data Parallelism}\label{subsec:cpdp_core}

CPDP introduces a synchronization period $K \ge 1$ that regulates
communication frequency. Numbering steps within each epoch as
$t=1,\ldots,T$, we define a synchronization indicator
\begin{equation}
s_t =
\begin{cases}
1, & \text{if } t \bmod K = 0 \text{ or } t=T, \\
0, & \text{otherwise.}
\end{cases}
\end{equation}

\paragraph{Local update phase}

If $s_t = 0$, each worker performs a local optimizer step, illustrated here using the simplified plain-SGD update:
\begin{equation}
\theta_{i,t+1} = \theta_{i,t} - \eta_t g_{i,t}.
\label{eq:local_update}
\end{equation}
During this phase, no gradient synchronization is performed,
and parameter trajectories may diverge across workers due to
independent local optimizer updates.

\paragraph{Synchronization phase}

When $s_t = 1$, CPDP executes a synchronization procedure composed of three stages. 
First, gradients are aggregated across workers via AllReduce and 
each worker applies its optimizer to the same averaged gradient; in simplified plain-SGD notation:
\begin{equation}
\widetilde{\theta}_{i,t+1}
=
\theta_{i,t} - \eta_t \bar{g}_t .
\end{equation}

Second, to reconcile the parameter divergence accumulated during local
updates, CPDP computes the average model across workers:
\begin{equation}
\theta^{\text{avg}}_{t+1}
=
\frac{1}{N}\sum_{i=1}^{N}
\widetilde{\theta}_{i,t+1}.
\label{eq:param_avg}
\end{equation}

Parameter averaging complements gradient synchronization because the two operations address different sources of divergence. The gradient AllReduce provides a shared gradient input at reconciliation steps, whereas parameter averaging reconciles the parameter differences accumulated during the preceding local-update interval. With local momentum buffers or adaptive moments, a shared gradient does not necessarily produce identical optimizer increments. The averaged parameters then define the reference used by the SlowMo correction. After the correction, all workers hold identical model parameters, while between synchronization boundaries they evolve locally.

Unlike classical LocalSGD, which performs only parameter averaging at synchronization boundaries, CPDP additionally performs a gradient
AllReduce step before parameter reconciliation. {This provides a shared gradient input at reconciliation boundaries while still reducing the frequency of gradient synchronization.
In the implementation evaluated in this study, optimizer internal states are not
synchronized. CPDP restores parameter consistency after each synchronization
boundary, but it does not enforce optimizer-state equivalence with DDP.}

\paragraph{Relation to the experimental optimizers}

More generally, let $q_{i,t}$ denote the local optimizer state of worker $i$ and $\mathcal{U}$ the configured optimizer update, including its weight-decay rule. A training step before parameter reconciliation is
\begin{equation}
(\widetilde{\theta}_{i,t+1},q_{i,t+1})=
\mathcal{U}(\theta_{i,t},q_{i,t},h_{i,t};\eta_t),
\end{equation}
where $h_{i,t}=g_{i,t}$ on local steps and $h_{i,t}=\bar g_t$ at CPDP synchronization boundaries. For momentum SGD, $q_{i,t}$ contains the momentum buffer; for AdamW, it contains the first- and second-moment estimates and step counter. The optimizer applies its own momentum, adaptive scaling, and weight-decay operations. On local steps, $\theta_{i,t+1}=\widetilde{\theta}_{i,t+1}$. At synchronization boundaries, the parameter average in Eq.~\eqref{eq:param_avg} is taken over the post-optimizer values $\widetilde{\theta}_{i,t+1}$, followed by the SlowMo correction below. Reconciliation changes parameters only: $q_{i,t+1}$ is neither averaged nor reset. SlowMo's $\beta$ is distinct from the SGD momentum coefficient and AdamW's internal moment coefficients. The simplified SGD equations describe communication placement, not the full experimental optimizer or a convergence result for momentum SGD or AdamW.

\subsection{{{SlowMo}} Momentum Stabilization}\label{subsec:slowmo}

To mitigate drift induced by reduced synchronization, CPDP integrates SlowMo momentum~\cite{wang_slowmo_2019} at each post-warmup synchronization boundary for $K>1$.

Let $\theta^{\text{ref}}_{t}$ denote the reference model stored after the previous synchronization boundary, and let $v_t$ denote the SlowMo velocity with coefficient $\beta \in [0,1)$. At synchronization:
\begin{align}
\Delta_t &= \theta^{\text{avg}}_{t+1} - \theta^{\text{ref}}_{t}, \\
v_{t+1} &\leftarrow \beta v_t + \Delta_t, \\
\theta^{\text{new}}_{t+1} &= \theta^{\text{ref}}_{t} + v_{t+1}, \\
\theta_{i,t+1} &\leftarrow \theta^{\text{new}}_{t+1} \quad \forall i, \\
\theta^{\text{ref}}_{t+1} &\leftarrow \theta^{\text{new}}_{t+1}.
\end{align}

The SlowMo update applies a momentum-filtered correction between the
newly averaged parameters and the previous synchronized reference
model. The velocity term $v_t$ accumulates these corrections over time,
acting as a low-pass filter that smooths abrupt parameter changes
between synchronization points.

The CPDP training procedure after warmup is summarized in
Algorithm \ref{alg:cpdp-slowmo}. The algorithm illustrates the
alternation between local update phases and synchronization phases,
as well as the integration of gradient aggregation, parameter averaging,
and SlowMo momentum correction at synchronization boundaries. During
warmup, the implementation uses the fully synchronous behavior described
in  Section~\ref{sec:implementation}, without periodic parameter
reconciliation or SlowMo correction.

\begin{algorithm}[htbp]
 \caption{CPDP with SlowMo momentum and a stateful local optimizer}
\label{alg:cpdp-slowmo}
\small
\begin{algorithmic}[1]
\REQUIRE Initial parameters $\theta$, synchronization period $K$, SlowMo coefficient $\beta$, learning-rate schedule $\eta_t$, optimizer $\mathcal{U}$
\STATE Initialize $\theta_i \leftarrow \theta$ for all workers
\STATE Initialize $v \leftarrow 0$, $\theta_{\text{ref}} \leftarrow \theta$
\STATE Initialize local optimizer state $q_i$ on each worker
\FOR{each epoch}
    \FOR{each step $t=1,\ldots,T$}
        \FOR{each worker $i$ \textbf{in parallel}}
            \STATE Sample mini-batch $\mathcal{B}_{i,t} \subset \mathcal{D}$ 
            \STATE Compute $g_{i,t} \leftarrow \nabla_\theta \ell(\theta_{i,t}; \mathcal{B}_{i,t})$
        \ENDFOR
        \IF{$t \bmod K \neq 0$ and $t\neq T$}
            \STATE $(\theta_{i,t+1},q_{i,t+1})\leftarrow\mathcal{U}(\theta_{i,t},q_{i,t},g_{i,t};\eta_t)$
        \ELSE
            \STATE $\bar{g}_t \leftarrow \frac{1}{N}\sum_i g_{i,t}$
            \STATE $(\widetilde\theta_{i,t+1},q_{i,t+1})\leftarrow\mathcal{U}(\theta_{i,t},q_{i,t},\bar g_t;\eta_t)$
            \STATE $\theta^{\text{avg}}_{t+1} \leftarrow \frac{1}{N}\sum_i \widetilde\theta_{i,t+1}$
            \STATE $v \leftarrow \beta v + (\theta^{\text{avg}}_{t+1} - \theta_{\text{ref}})$
            \STATE $\theta_{\text{new}} \leftarrow \theta_{\text{ref}} + v$
            \STATE $\theta_{i,t+1} \leftarrow \theta_{\text{new}}$; retain $q_{i,t+1}$ locally
            \STATE $\theta_{\text{ref}} \leftarrow \theta_{\text{new}}$
        \ENDIF
    \ENDFOR
\ENDFOR
\RETURN $\theta_{\text{ref}}$
\end{algorithmic}
\end{algorithm}

\subsection{Communication Complexity}\label{subsec:cpdp_summary}

Let $T$ denote the number of mini-batch steps per epoch. 
Table~\ref{tab:comm_comparison} compares the collective communication 
operations per epoch across the three methods evaluated in this work.

\begin{table}[htbp]
\centering
\caption{Communication operations per epoch for each method.}
\label{tab:comm_comparison}
\small
\resizebox{\columnwidth}{!}{%
\begin{tabular}{lcc}
\toprule
{Method} & {Gradient AllReduce} & {Parameter AllReduce} \\
\midrule
DDP ($K=1$) & $T$ & 0 \\
LocalSGD ($K$) & 0 & $\lceil T/K \rceil$ \\
CPDP ($K$) & $\lceil T/K \rceil$ & $\lceil T/K \rceil$ 
\\
\bottomrule
\end{tabular}%
}
\end{table}

The counts in Table~\ref{tab:comm_comparison} apply to post-warmup epochs. DDP performs one gradient AllReduce per mini-batch. LocalSGD removes per-step gradient synchronization and averages parameters every $K$ steps. CPDP performs both a gradient AllReduce and a parameter AllReduce at each synchronization boundary. Thus, CPDP is more expensive than LocalSGD per synchronization event, but it provides a synchronized gradient signal before parameter reconciliation, which parameter averaging alone does not provide. This cost--benefit trade-off is evaluated empirically in  Section~\ref{subsec:hidden_comm}, where exposed synchronization time is profiled directly. In our implementation, optimizer states remain local between synchronization boundaries. During warmup, CPDP and LocalSGD temporarily use $K=1$,
with per-step gradient synchronization and without periodic parameter
reconciliation; all reported wall-clock times include this warmup cost.

\section{Implementation}\label{sec:implementation}

This section describes the implementation of CPDP within a standard distributed deep learning framework. The implementation is designed to integrate with existing distributed training pipelines without modifying model architectures, loss functions, or communication libraries.

\subsection{Distributed Execution Environment}

CPDP is implemented as an execution-layer extension over PyTorch DDP using the NCCL backend for inter-GPU communication. Each training process is bound to a single GPU and participates in a global process group initialized through environment-variable-based distributed 
launch. The implementation supports both single-node multi-GPU and multi-node deployments over Ethernet interconnects. No modifications to NCCL primitives, collective algorithms, or backend communication libraries are introduced.

Training data are partitioned using PyTorch's 
\texttt{DistributedSampler} to ensure disjoint mini-batches across workers. For intra-site experiments, synchronized Batch Normalization is employed to maintain consistent statistics across workers. {For the cross-site WAN experiments, synchronized Batch Normalization is disabled and per-worker BatchNorm statistics are used instead, to avoid adding an additional batch-level synchronization operation over the high-latency WAN link. This setting matches the cross-site configuration used in the reported experiments.} Evaluation is performed without distributed 
sampling to ensure correct global accuracy measurement.

For cross-site experiments, the distributed process group spans nodes located at different Grid'5000 sites, connected over the WAN. NCCL communication in this configuration uses TCP sockets over the site-to-site network link. No special configuration of NCCL transport or routing is applied; 
the default NCCL socket-based backend is used for all inter-node communication in both intra-site and cross-site deployments.

\subsection{Synchronization Control Mechanism}

The central mechanism of CPDP is the explicit control of gradient synchronization frequency during backpropagation. Given synchronization period $K$, gradient AllReduce is enabled only every $K$ mini-batch steps, or at the final step of an epoch to ensure that model replicas remain synchronized before the next epoch begins.

This behavior is implemented using DDP's \texttt{no\_sync()} context manager on non-synchronization steps, which suppresses DDP's gradient reduction hooks while preserving local gradient computation. On synchronization steps, \texttt{nullcontext()} is used, allowing DDP's
default NCCL-based AllReduce to execute during the backward pass.

At each training step, CPDP follows a fixed execution order. First,
\texttt{loss.backward()} is called under the appropriate synchronization
context. On synchronization steps, DDP's gradient AllReduce is triggered during
the backward pass, producing averaged gradients on all workers. On local steps,
the \texttt{no\_sync()} context suppresses gradient communication.
Second, \texttt{optimizer.step()} is applied. On synchronization steps, the
optimizer update uses the averaged gradients produced by DDP. On local steps,
each worker updates its parameters using its own local gradients. Third, outside the warmup phase, CPDP calls \texttt{slowmo.synchronize()} at synchronization boundaries. This operation averages model parameters across
workers, applies the SlowMo correction, and writes the reconciled values back to
the model parameters using \texttt{param.data.copy\_()}. This step modifies only
model parameter tensors. It does not modify optimizer internal states such as
SGD momentum buffers or AdamW first- and second-moment estimates.

\paragraph{Implication for Optimizer State}

Optimizer internal states therefore remain local between synchronization boundaries. CPDP restores equality of model parameters after each reconciliation step, but it does not enforce optimizer-state equivalence with standard DDP. The method should therefore be interpreted as a parameter-reconciliation strategy with local optimizer state, not as an exact replication of the DDP optimizer trajectory.

This distinction is relevant for both SGD and AdamW experiments. In the ResNet experiments, SGD momentum buffers are local. In the ViT-S experiments, AdamW moment estimates are also local. Explicit optimizer-state reconciliation is a
possible implementation variant, but it was not used in the experiments reported in this study.

Importantly, CPDP operates purely at the {parameter} synchronization
layer. It does not modify optimizer update rules, model architectures,   {loss functions,} or {NCCL communication primitives}.

\subsection{Training Protocol}

All experiments employ cosine annealing learning-rate scheduling with linear warmup.
For intra-site ResNet-50 scaling experiments, the base learning rate is scaled
linearly with global batch size following standard practice, from $\eta_0=0.1$ at
global batch size 256 up to $\eta_0=0.4$ at global batch size 1024. For the main
cross-site WAN experiment, we use a fixed base learning rate of $\eta_0=0.1$ for all
three synchronization strategies and report $\eta_0=0.4$ separately as a
learning-rate sensitivity setting. During warmup epochs, CPDP and LocalSGD use $K=1$.
Consequently, gradient AllReduce is performed at every step, while
the periodic parameter-reconciliation and SlowMo operations are not
applied. After warmup, the configured synchronization period is restored.
Learning-rate updates are applied consistently across all workers.

The same codebase is used across all synchronization configurations,
with method-specific behavior determined by the synchronization strategy
and the synchronization period $K$. To ensure deterministic behavior, random seeds are initialized deterministically while incorporating the process rank to ensure reproducibility across runs while avoiding identical sampling across workers, and non-deterministic   CUDA optimizations are disabled. This design limits auxiliary implementation differences, while the evaluated synchronization strategies still differ in gradient aggregation, parameter averaging, and SlowMo correction.

\subsection{Instrumentation}

The training loop is instrumented to record per-epoch metrics including training loss, training accuracy, test accuracy, wall-clock epoch time, and throughput (images per second). Epoch time is further decomposed into four components: forward pass time ($T_{\text{fwd}}$), backward pass time ($T_{\text{bwd}}$), explicit parameter 
synchronization time ($T_{\text{comm}}$), and residual overhead.

For DDP, $T_{\text{comm}}$ is not measured explicitly because gradient AllReduce operations are overlapped with the backward pass by the NCCL backend. As a result, communication time is partially hidden within the backward computation phase. For CPDP and LocalSGD, $T_{\text{comm}}$ captures the explicit parameter AllReduce performed outside the backward pass; for CPDP, it additionally includes the SlowMo correction. This decomposition enables the estimation of {DDP's backward-time inflation proxy} as discussed in  Section~\ref{sec:results}.

In addition, the number of gradient synchronization events and parameter reconciliation operations are tracked per epoch. All metrics are logged in structured CSV format with per-epoch granularity to support reproducibility and post-hoc analysis.

\section{Experimental Setup}\label{sec:experimental_setup}

This section describes the experimental setup used to evaluate CPDP, including the hardware platform, training workloads, synchronization configurations, and evaluation metrics. Within each comparison, methods share the stated optimizer settings and system conditions. Their synchronization and reconciliation rules differ, and the comparisons across deployment regimes do not isolate network latency alone.

\subsection{Experimental Platform}

Experiments are conducted on the Grid'5000 testbed, a large-scale distributed systems research infrastructure in France~\cite{grid5000}.

\paragraph{Intra-site cluster ({{Nancy}})}

The \emph{graffiti} cluster provides GPU-enabled nodes used for
all intra-site scaling experiments. Each node contains two Intel
Xeon Silver 4110 CPUs (8 cores per CPU), four NVIDIA GeForce RTX
2080 Ti GPUs, 128\,GB of system memory, and a 10\,Gb Ethernet
interconnect. Intra-site experiments scale from 4~GPUs (1 node)
to 16 GPUs (4 nodes), with inter-node communication over the
10\,Gb Ethernet network.

\paragraph{Cross-site deployment (Nancy $\leftrightarrow$ Sophia)}

Cross-site experiments deploy 4 GPUs on a \emph{graffiti} node in
Nancy and 4 GPUs on an \emph{esterel} node in Sophia, for a total
of 8~GPUs across two geographic sites. The \emph{esterel} node
provides identical CPUs and GPUs (RTX 2080 Ti), 96\,GB of system
memory, a 1\,Gb Ethernet interface, and a 40\,Gb InfiniBand
adapter. In our experiments, inter-site communication occurs
over the Ethernet network using NCCL's TCP socket transport,
without exploiting InfiniBand acceleration. The measured
round-trip latency between sites is approximately 16.6\,ms. {The two sites also differ in local interface bandwidth (10\,Gb Ethernet at Nancy, 1\,Gb Ethernet at Sophia), so the effective cross-site throughput is bounded by the slower 1\,Gb link and the inter-site path. This asymmetry is part of the realistic WAN setting we evaluate; all three synchronization strategies are compared under identical network conditions, while its performance impact can differ with the synchronization strategy.}

Nodes are reserved in exclusive mode to avoid interference from co-located workloads. The software stack consists of 
Python~3.9, PyTorch~2.4, CUDA~11.8, and NCCL~2.18. Distributed 
training uses PyTorch DistributedDataParallel with NCCL 
collectives. No custom communication primitives or backend 
modifications are introduced.

To ensure reproducibility, random seeds are fixed for each run using three independent seeds (42, 123, 777). CUDA deterministic execution is enabled and distributed process initialization is synchronized across workers. Each configuration is executed three times using the different seeds, and reported results correspond to the mean and standard deviation across these runs.

No mixed-precision training, gradient compression, gradient 
sparsification, model sharding, or memory-saving techniques 
are used. The reported results compare the complete synchronization strategies under the stated protocols.

\subsection{Training Workloads}

We evaluate synchronization control across four configurations designed to test different aspects of the communication--accuracy trade-off.

\paragraph{{{ResNet}}-50 on {{CIFAR}}-100 (Intra-site Strong Scaling)}

We train ResNet-50~\cite{he_deep_2015} on CIFAR-100 using SGD 
with momentum (0.9) and weight decay ($5 \times 10^{-4}$). This 
workload is used for the primary scaling study from 4 to 16 GPUs. 
The per-GPU batch size is fixed at 64, yielding global batch sizes 
from 256 (4 GPUs) to 1024 (16 GPUs). The learning rate is scaled 
linearly with global batch size (base $\eta_0=0.1$ at batch size 
256). Training runs for 100~epochs with 5 warmup epochs and 
cosine annealing.

\paragraph{{{ResNet}}-50 on {{CIFAR}}-100 (Cross-site {{WAN}})}

The same ResNet-50 architecture is trained across the Nancy--Sophia WAN link on 8~GPUs (4 per site). For this configuration, the per-GPU batch size is 128, yielding a global batch size of 1024.
{The main cross-site WAN experiment uses a fixed base learning rate of $\eta_0=0.1$ for all three synchronization strategies. This configuration is reported as a stable WAN training setting rather than as a linear learning-rate scaling experiment.}
{To characterize aggressive learning-rate scaling in this regime, we additionally report a cross-site $\eta_0=0.4$ diagnostic over three seeds, using the main $\beta=0.3$ configuration and applying the same linearly scaled learning rate identically to all three methods. Under this setting, DDP ($K=1$) reaches $77.39\pm0.70\%$, LocalSGD ($K=2$) reaches $77.31\pm1.04\%$, and CPDP ($K=2$) reaches $76.65\pm0.93\%$. The three methods produce close peak accuracies relative to the observed inter-seed variability, with DDP marginally ahead. This contrasts with the fixed $\eta_0=0.1$ WAN protocol, under which CPDP improves over both baselines. We therefore interpret CPDP as improving the accuracy--time trade-off under stable synchronization configurations rather than universally dominating DDP across all learning-rate regimes, and we adopt $\eta_0=0.1$ as the main WAN comparison while reporting $\eta_0=0.4$ as a sensitivity result. The aggressive scaling also increases inter-run variability relative to the $\eta_0=0.1$ protocol; for example, CPDP increases from ${\pm}0.11\%$ at $\eta_0=0.1$ to ${\pm}0.93\%$ at $\eta_0=0.4$, consistent with the drift mechanism documented in  Section~\ref{subsec:lr_sensitivity}.}

\paragraph{ResNet-50 on {TinyImageNet} (Dataset Generality)}

To verify generalization beyond CIFAR-100, we train ResNet-50 on 
TinyImageNet (200 classes, 64~\ensuremath{\times}~64 resolution) at two scales: 
4~GPUs (1~node) and 12 GPUs (3 nodes). Hyperparameters follow the 
CIFAR-100 configuration with appropriate adjustments for the 
larger dataset.

\paragraph{{{ViT}}-{S} on {{CIFAR}}-100 (Architecture Generality)}

To evaluate behavior on transformer architectures, we train 
ViT-S~\cite{dosovitskiy_image_2021} on CIFAR-100 using AdamW 
with base learning rate $\eta_0=5\times10^{-4}$ and weight decay 0.05. 
Training runs for 200 epochs at three scales (4, 6, and 8~GPUs). 
This workload exhibits a higher compute-to-communication ratio 
than ResNet-50, allowing assessment of synchronization control 
when computation dominates iteration time.

\subsection{Training Configuration}

Table~\ref{tab:hyperparameters} summarizes the optimizer settings and training hyperparameters used for each workload. All synchronization strategies (DDP, CPDP, and LocalSGD) share identical optimizer configurations, batch sizes, and learning rate schedules at each scale. Only the synchronization strategy differs across configurations.

\begin{table}[htbp]
\centering
\caption{Hyperparameters for each training workload.}
\label{tab:hyperparameters}
\small
\setlength{\tabcolsep}{2.5pt}
\resizebox{\columnwidth}{!}{%
\begin{tabular}{lcc}
\toprule
{Parameter} & {ResNet-50} & {ViT-S} \\
\midrule
Dataset & CIFAR-100/TinyImageNet & CIFAR-100 \\
Batch size (per GPU) & 64 (intra)/128 (cross) & 64 \\
Base learning rate $\eta_0$ & 0.1 & $5 \times 10^{-4}$ \\
Learning-rate protocol &
\makecell{Linear intra-site;\\
fixed $\eta_0=0.1$ cross-site;\\
$\eta_0=0.4$ sensitivity} &
Fixed base rate \\
Optimizer & SGD (mom.\ 0.9) & AdamW \\
Weight decay & $5 \times 10^{-4}$ & 0.05 \\
Epochs & 100 & 200 \\
Warmup epochs & 5 (CIFAR-100) /  10 (TinyImageNet) & 20 \\
Learning-rate schedule & Cosine & Cosine \\
SlowMo coefficient $\beta$ & 0.3 & 0.3 \\
Synchronization period $K$ & 1, 2, 4, 8 & 1, 2 \\
Seeds & \multicolumn{2}{c}{42, 123, 777} 
\\
\bottomrule
\end{tabular}%
}
\end{table}

{The value $\beta=0.3$ is used as the fixed CPDP configuration in the main experiments. Warmup is fixed within each workload and shared by all synchronization strategies compared in the same table. We use 5 warmup epochs for ResNet-50/CIFAR-100, 10 warmup epochs for ResNet-50/TinyImageNet, and 20 warmup epochs for ViT-S/CIFAR-100. As the SlowMo coefficient sensitivity analysis in  Section~\ref{subsec:k_sweep} shows, $\beta=0.3$ attains the highest
mean accuracy within the stable range, while overly large values destabilize training.}

The synchronization period $K$ is the interval in mini-batch steps between regular synchronization boundaries. There are up to $K-1$ local steps before each boundary; the last step of an epoch also synchronizes. The DDP baseline uses $K=1$ and no SlowMo correction.

Table~\ref{tab:scaling_config} details the GPU scaling configurations used in the intra-site scaling study. The per-GPU batch size is fixed at 64, and the learning rate is scaled linearly with the global batch size following standard large-batch training practice.

\begin{table}[htbp]
\centering
\caption{Intra-site scaling configurations for ResNet-50/CIFAR-100.}
\label{tab:scaling_config}
\small
\begin{tabular}{ccccc}
\toprule

{Label} & {GPUs} & {Nodes} 
& {Global BS} & $\eta_0$ \\
\midrule
g4  & 4  & 1 & 256  & 0.10 \\
g6  & 6  & 2 & 384  & 0.15 \\
g8  & 8  & 2 & 512  & 0.20 \\
g12 & 12 & 3 & 768  & 0.30 \\
g16 & 16 & 4 & 1024 & 0.40 
\\
\bottomrule
\end{tabular}
\end{table}

\subsection{Synchronization Strategies}

We compare three synchronization strategies. The first baseline is
standard synchronous data parallelism (DDP), corresponding to a
synchronization period $K=1$, where gradient AllReduce is performed
at every mini-batch step and no parameter averaging is applied.

The second baseline is LocalSGD, where workers perform local updates between synchronization points and periodically average parameters across workers using AllReduce, without gradient synchronization at the averaging step and without
SlowMo correction (the configured local optimizer still uses its own momentum or adaptive moments). The primary comparison uses $K=2$, and additional synchronization-period sensitivity experiments evaluate $K=4$ and $K=8$ where
specified.

The third method is CPDP, where workers also perform local updates between synchronization points, but at synchronization steps gradients are first aggregated via AllReduce, followed by parameter averaging and SlowMo momentum
correction. The primary comparison uses $K=2$, with additional sensitivity
experiments at larger synchronization periods.

More generally, synchronization occurs whenever $t \bmod K = 0$ or at the final step $t=T$ of an epoch. Workers therefore perform up to $K-1$ local updates between two
synchronization steps. When $K=1$, synchronization occurs at every
iteration. This specifies the communication frequency, while the DDP baseline additionally omits parameter reconciliation and SlowMo correction.

During warmup epochs, both CPDP and LocalSGD temporarily use
$K=1$, so gradient synchronization is performed at every step and
their periodic parameter-reconciliation operations are not applied.
After warmup, the configured synchronization period is restored.
All other hyperparameters remain identical across the three methods.

For the synchronization period ablation ($K$-sweep), we additionally evaluate
CPDP and LocalSGD at $K \in \{2,4,8,16,32,64\}$ on the 8-GPU configuration.
Values with $K\geq 16$ are treated as exploratory boundary-condition runs
rather than recommended operating points.

\subsection{Evaluation Methodology}

Performance is evaluated using several complementary metrics. Model quality is measured using test accuracy on the held-out dataset, averaged across three independent runs. {Unless otherwise stated, reported test accuracy refers to the peak (best) test accuracy over the 100-epoch run. For the main accuracy comparisons, we additionally report statistical significance using Welch's two-sided $t$-test over the three random seeds, computed on this same peak test-accuracy metric. Given the limited number of seeds, we interpret $p$-values alongside the reported inter-run variability and treat borderline cases conservatively.} System performance is evaluated using the total wall-clock training time across all epochs and the achieved throughput measured in processed images per second. To analyze the sources of execution time, we decompose epoch duration into forward pass time ($T_{\text{fwd}}$), backward pass time ($T_{\text{bwd}}$), explicit synchronization time ($T_{\text{comm}}$), and residual overhead. In addition, we record the number of gradient synchronization and parameter reconciliation events per epoch. This decomposition enables the estimation of {DDP's backward-time inflation} as discussed in  Section~\ref{sec:results}. All metrics are aggregated using collective operations to ensure correctness, and accuracy evaluation is performed in inference mode without gradient computation.

To ensure a fair comparison, all methods use identical model architectures, datasets, and hyperparameters at each scale, and only the synchronization strategy differs between configurations. Each experiment is executed for the same number of epochs, and results are reported as the mean and standard deviation across three independent runs. {For intra-site scaling experiments, linear learning-rate scaling with respect to global batch size is applied uniformly across all methods. For the
main cross-site WAN experiment, a fixed base learning rate is used for all synchronization strategies to provide a stable WAN training configuration. In
each configuration, DDP, LocalSGD, and CPDP use the same optimizer settings, batch size, learning-rate schedule, and number of epochs, so no synchronization strategy benefits from method-specific tuning.} These comparisons evaluate each synchronization strategy as a whole. They do not separately identify the effects of WAN latency, gradient aggregation, parameter averaging, and SlowMo correction.

\section{Results and Discussion}\label{sec:results}

This section presents the evaluation of CPDP across intra-site scaling, cross-site WAN training, and multiple architectures and datasets. We compare three synchronization strategies: DDP, LocalSGD, and CPDP under identical conditions at each scale. We first present the cross-site WAN results, then analyze DDP's backward-time inflation proxy, followed by intra-site scaling, generality experiments, and the synchronization period ablation.

\subsection{Cross-site {{WAN}} Training}\label{subsec:cross_site}

The cross-site configuration represents the most communication-constrained setting in our evaluation: 8~GPUs split across Nancy (graffiti, 10\,GbE) and Sophia (esterel, 1\,GbE), with approximately 16.6\,ms round-trip latency between sites. Hereafter, ``pp'' denotes percentage points. Table~\ref{tab:cross_site} presents the results.

\begin{table}[htbp]
\centering
\caption{Cross-site WAN training. ResNet-50/CIFAR-100, 8~GPUs
(4 Nancy + 4 Sophia), 100~epochs. Accuracy values report mean ${\pm}$ standard
deviation of peak test accuracy over 3 seeds; time values report mean $\pm$
standard deviation of total wall-clock training time.}
\label{tab:cross_site}
\resizebox{\columnwidth}{!}{%
\begin{tabular}{lcccc}
\toprule
{Method} & {$K$} & {Peak Acc.\ (\%)} & {Time (s)} & {vs DDP (pp)} \\
\midrule
DDP & 1 & 73.84~\ensuremath{\pm}~0.23 & 4959~\ensuremath{\pm}~65 & -- \\
\hline
LocalSGD & 2 & 74.12~\ensuremath{\pm}~0.75 & 6922~\ensuremath{\pm}~65 & \ensuremath{+}0.28 \\
LocalSGD & 4 & 74.51~\ensuremath{\pm}~0.22 & 4170~\ensuremath{\pm}~41 & \ensuremath{+}0.67 \\
\hline
CPDP & 2 & 76.12~\ensuremath{\pm}~0.11 & 7203~\ensuremath{\pm}~129 & \ensuremath{+}2.28 \\
\textbf{CPDP} & \textbf{4} & \textbf{76.28~\ensuremath{\pm}~0.34} & \textbf{4275~\ensuremath{\pm}~59} & $\mathbf{+2.44}$ 
\\
\bottomrule
\end{tabular}%
}
\end{table}

{ Under the fixed $\eta_0=0.1$ WAN protocol, CPDP improves over both baselines at both evaluated synchronization periods. At $K=2$, CPDP attains 76.12\% (\ensuremath{+}2.28 percentage points over DDP) with the lowest inter-run standard deviation. At $K=4$, CPDP attains 76.28\% (\ensuremath{+}2.44 percentage points over DDP and \ensuremath{+}1.77 percentage points over LocalSGD at the same synchronization period) while also reducing average wall-clock time by approximately 13.8\% relative to DDP (4275~s versus 4959~s on average). The two CPDP settings are not statistically different in accuracy (Welch's two-sided $t$-test, $p=0.50$), so $K=4$ should not be read as more accurate than $K=2$; its advantage is that it reaches the same accuracy level while reducing average training time below DDP. Thus, within the cross-site regime, increasing the synchronization period from $K=2$ to $K=4$ preserves CPDP's accuracy advantage while turning its wall-clock cost into a wall-clock saving.}

However, the per-step global synchronization barrier in DDP,
combined with the high latency of the cross-site WAN link, exposes
each iteration to network latency and synchronization variability.
{As discussed in  Section~\ref{subsec:hidden_comm},
DDP's backward time increases substantially in the cross-site setting
because gradient AllReduce is launched during the backward pass and
overlapped with computation. This increase should not be interpreted
as pure blocked communication time, but it indicates that DDP's
backward phase is strongly affected by WAN communication.}
These timing measurements characterize execution cost; they do not establish that WAN latency causes the observed accuracy differences.

After warmup, CPDP performs gradient synchronization approximately every $K$ steps, followed by parameter reconciliation, with an additional boundary at the end of each epoch when needed. This changes communication cost and the optimization trajectory. The higher peak test accuracy observed here is conditional on the chosen learning-rate schedule, $K$, and SlowMo coefficient $\beta$; it cannot be attributed to reduced WAN exposure alone. The sensitivity results in  Sections \ref{sec:experimental_setup}, \ref{subsec:lr_sensitivity}{, and } \ref{subsec:k_sweep} demonstrate that these optimization choices affect the outcome. {Welch's two-sided $t$-test over three seeds, computed on the same peak test-accuracy metric reported in the table, gives $p=0.0008$ for CPDP $K=2$ vs.\ DDP and $p=0.042$ for CPDP $K=2$ vs.\ LocalSGD $K=2$. Both differences are statistically significant at the 0.05 level, though the latter is interpreted conservatively given the small number of seeds.} {At $K=4$, CPDP improves over DDP with $p=0.0010$ and over LocalSGD at the same synchronization period with $p=0.0031$, both significant at the 0.05 level.}

\begin{figure}[htbp]
\centering
\includegraphics[width=\linewidth]{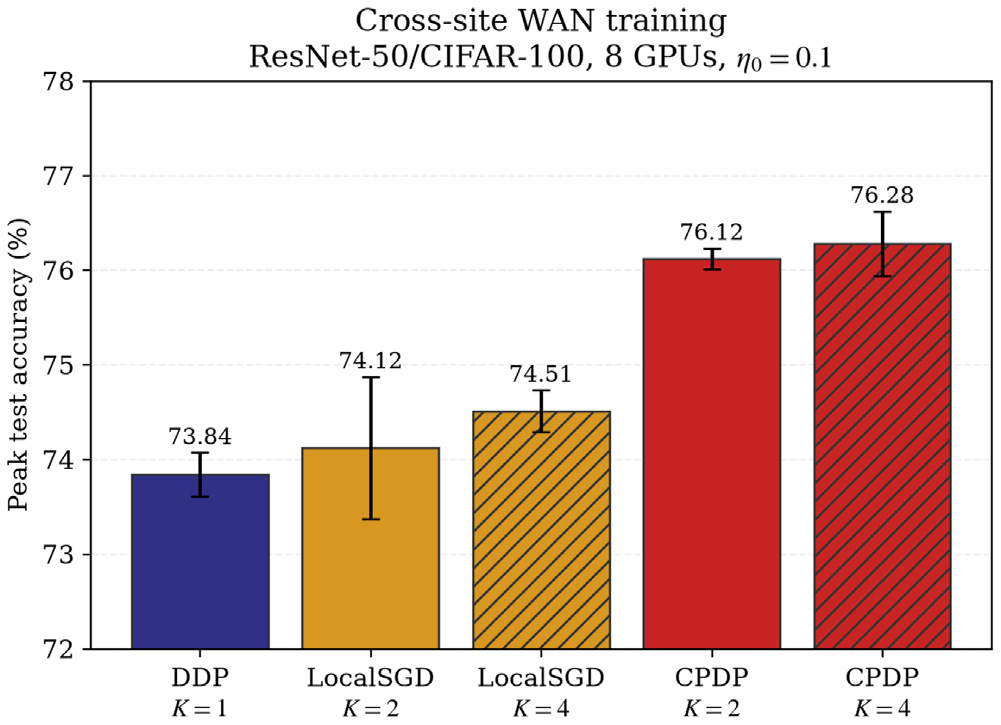}
\caption{Cross-site WAN training on ResNet-50/CIFAR-100 with 8~GPUs distributed across Nancy and Sophia (4+4 GPUs). Bars show mean peak test accuracy over
3 seeds; error bars denote one standard deviation. Solid bars correspond to DDP ($K=1$) and the conservative periodic setting ($K=2$), while hatched
bars correspond to $K=4$. Under the fixed $\eta_0=0.1$ WAN protocol, CPDP improves over DDP at both synchronization periods (\ensuremath{+}2.28 percentage points at $K=2$, \ensuremath{+}2.44 percentage points at $K=4$); the two CPDP settings are not statistically different in accuracy, but $K=4$ additionally reduces average wall-clock time below DDP.}
\label{fig:cross_site_accuracy}
\end{figure}

Figure~\ref{fig:cross_site_accuracy} visualizes these {peak test accuracy} results under the fixed $\eta_0=0.1$ cross-site configuration.

To quantify the effect at the conservative $K=2$ periodic setting, Table~\ref{tab:cross_site_degradation} compares each method's cross-site accuracy against its intra-site accuracy at the same GPU count (8~GPUs).
{The two settings correspond to different deployment regimes by design: the intra-site point follows the linear-scaling protocol of the
strong-scaling study, whereas the cross-site point uses the stable WAN learning rate. The comparison therefore characterizes end-to-end degradation between two
realistic operating conditions rather than isolating WAN latency as a single controlled variable.}

\begin{table}[htbp]
\centering
\caption{Accuracy degradation from intra-site to cross-site training at the conservative $K=2$ periodic setting (8~GPUs, ResNet-50/CIFAR-100).
Mean $\pm$ standard deviation of peak test accuracy over 3 seeds. The two settings use different learning-rate protocols and global batch sizes, so the comparison reflects end-to-end change across deployment regimes.}
\label{tab:cross_site_degradation}
\small
\begin{tabular}{lccc}
\toprule

{Method} & {Intra (\%)} 
& {Cross (\%)} & {$\Delta$} \\
\midrule
DDP & 78.94~\ensuremath{\pm}~0.50 & 73.84~\ensuremath{\pm}~0.23 & \ensuremath{-}5.10 \\
LocalSGD & 77.36~\ensuremath{\pm}~0.19 & 74.12~\ensuremath{\pm}~0.75 & \ensuremath{-}3.24 \\
\textbf{CPDP} & 77.97~\ensuremath{\pm}~0.44 & \textbf{76.12~\ensuremath{\pm}~0.11}
& $\mathbf{-1.85}$ 
\\
\bottomrule
\end{tabular}
\end{table}

DDP suffers the largest degradation (\ensuremath{-}5.10 percentage points) when moving from intra-site to cross-site training, whereas CPDP exhibits the smallest degradation (\ensuremath{-}1.85 percentage points), retaining most of its intra-site accuracy. At the conservative $K=2$ periodic setting, CPDP therefore demonstrates the smallest degradation across the evaluated deployment regimes.

\subsection{{Communication Pressure and Direct Profiling}
}\label{subsec:hidden_comm}

{In PyTorch DDP, gradient AllReduce is launched through autograd hooks and overlapped with the backward pass via NCCL bucketed communication. As
a result, DDP does not expose a separate communication-time term in simple wall-clock instrumentation. The portion of time that is truly blocked on communication cannot be measured from epoch timers alone; it would require CUDA/NCCL trace-level profiling.
The CPDP timing used in this proxy corresponds to the $K=2$ cross-site configuration, so the proxy should be interpreted for the conservative periodic setting rather than for the $K=4$ configuration.}

{We therefore report a backward-time inflation proxy rather than an exact blocked-communication measurement. For DDP, $T_{\text{bwd}}$ includes backward computation, overlapped gradient communication, and any synchronization waiting exposed during the backward pass. For CPDP, gradient AllReduce is suppressed during local steps using \texttt{no\_sync()}, so its measured backward time corresponds to the periodic execution path without per-step gradient synchronization. However, this should not be interpreted as a pure-compute baseline. The CPDP backward path is not identical to the DDP backward path because DDP communication hooks are suppressed during local steps and PyTorch's execution context differs under \texttt{no\_sync()}. We define the
backward-time inflation proxy as:}
\begin{equation}
{\Delta T_{\text{bwd}}} =
T_{\text{bwd}}^{\text{DDP}} -
T_{\text{bwd}}^{\text{CPDP}}
\label{eq:hidden_proxy}
\end{equation}

{This proxy should not be interpreted as pure communication time or as the fraction of time blocked on the WAN link. It measures the increase in observed backward-pass time under DDP relative to the periodic execution path used by CPDP. The value may include overlapped gradient communication, synchronization waiting, and execution-  context differences introduced by DDP
communication hooks and the use of \texttt{no\_sync()}. We use it only as an
indicator of communication pressure inside DDP's backward phase, not as a direct decomposition of communication and computation.}

\begin{table}[t]
\centering
\caption{{DDP backward-time inflation proxy (per epoch).
$T_{\text{bwd}}$ values are averaged over the last three steady-state epochs
from a matched representative run. All intra-site rows were measured using a
matched node allocation with verified-consistent forward-pass times across
methods. The ``\% Epoch'' column reports $\Delta T_{\text{bwd}}$ relative to the
corresponding DDP epoch time. The proxy is reported as an indirect indicator of
DDP backward-phase inflation, whereas Tables~\ref{tab:cross_comm_profile}--\ref{tab:intra_comm_profile}
report direct exposed-synchronization measurements.}}
\label{tab:hidden_comm}
\small
\begin{tabular}{lcccc}
\toprule

{Scale} & {$T_{\text{bwd}}^{\text{DDP}}$}
& {$T_{\text{bwd}}^{\text{CPDP}}$}
& {{$\Delta T_{\text{bwd}}$}}
& {\% Epoch} \\
\midrule
{g4}  & {17.5~s} & {15.5~s} & {2.0~s} & {6.9\%} \\
{g8}  & {16.5~s} & {12.1~s} & {4.4~s} & {18.9\%} \\
{g12} & {15.1~s} & {10.4~s} & {4.7~s} & {23.6\%} \\
{g16} & {11.5~s} & {7.8~s}  & {3.7~s} & {23.6\%} \\
{g8 (cross)} & {46.8~s} & {26.3~s} & {20.5~s} & {41.6\%} 
\\
\bottomrule
\end{tabular}
\end{table}

{
Table~\ref{tab:hidden_comm} shows that DDP's backward-time inflation proxy
increases from 6.9\% at g4 to 18.9\% at g8 and reaches 23.6\% at g12--g16.
The plateau between g12 and g16 suggests that intra-site communication pressure
saturates under this 10\,Gb Ethernet deployment. In contrast, the cross-site WAN
configuration reaches 41.6\%, indicating substantially stronger communication
pressure inside DDP's overlapped backward phase. The intra-site and cross-site
rows correspond to different deployment regimes, so the increase in absolute
backward time across these regimes cannot be attributed solely to the network: the deployment protocols also differ in batch size and other settings. As discussed above, this proxy should not be
interpreted as pure blocked communication time; it is used only as an indicator
of the extent to which DDP's backward phase is affected by communication and
synchronization behavior.}

{The previous observation helps interpret execution time, not the cause of the accuracy differences in  Section~\ref{subsec:cross_site}. CPDP reduces gradient synchronization frequency at the cost of explicit parameter reconciliation. Neither the backward-time proxy nor the direct profiling isolates the contribution of WAN latency to optimization quality.}

{
The backward-time inflation proxy above is useful for comparing communication pressure across scales, but it remains an indirect estimate because it is derived from differences in observed backward-pass time rather than from direct profiling of synchronization events. To complement this proxy, we therefore add direct exposed-synchronization profiling experiments. We first report the cross-site WAN profiler, where communication exposure is most critical, and then report the intra-site profiler for representative scales.}

{
This profiler measures the exposed time associated with collective
synchronization events under the same cross-site deployment as
Table~\ref{tab:cross_site}. In contrast to the backward-time proxy in
Table~\ref{tab:hidden_comm}, this measurement is not based on subtracting
DDP and CPDP backward-pass durations. It therefore avoids the main limitation of
the proxy, namely that \texttt{no\_sync()} changes the autograd execution context
during local steps. We use the profiler to quantify exposed synchronization
time directly and retain the proxy only as a scale-wise indicator of
communication pressure.}

\begin{table}[t]
\centering
\caption{{Direct cross-site profiling of exposed
synchronization time on ResNet-50/CIFAR-100 with 8~GPUs
(4 Nancy + 4 Sophia), averaged over three seeds. ``Syncs/epoch'' is the number
of synchronization events per epoch; ``Per-sync'' is the mean exposed time per
synchronization event; ``Exposed/epoch'' is their product; and ``Compute floor''
is the measured communication-free compute baseline.}}
\label{tab:cross_comm_profile}
\small
\resizebox{\columnwidth}{!}{%
\begin{tabular}{lccccc}
\toprule

{Method} & {$K$} & {Syncs/epoch} &
{Per-sync} & {Exposed/epoch} & {Compute floor} \\
\midrule
DDP      & 1 & 48 & 0.83~s & 40.0~s & 8.9~s \\
\hline
LocalSGD & 2 & 24 & 1.66~s & 39.9~s & 8.9~s \\
CPDP     & 2 & 24 & 1.66~s & 40.0~s & 8.8~s \\
\hline
LocalSGD & 4 & 12 & 1.71~s & 20.5~s & 8.9~s \\
\textbf{CPDP} & \textbf{4} & 12 & 1.70~s & $\mathbf{20.5}$~s & 9.0~s 
\\
\bottomrule
\end{tabular}%
}
\end{table}

{
Table~\ref{tab:cross_comm_profile} confirms that the compute floor is nearly identical across methods (8.8--9.0~s per epoch), so the observed differences are driven primarily by synchronization behavior. DDP performs 48 synchronization events per epoch and exposes about 40.0~s of synchronization time. At $K=2$, CPDP and LocalSGD perform only 24 synchronization events, but each event is roughly twice as expensive as a DDP event; consequently, their total exposed synchronization time remains comparable to DDP.

This should be interpreted together with the operation counts in Table~\ref{tab:comm_comparison}. At $K=2$, CPDP performs $\lceil T/2\rceil$ gradient AllReduce operations and $\lceil T/2\rceil$ parameter AllReduce operations, giving the same order of collective-operation count as DDP's $T$ per-step gradient AllReduce operations, but with a different placement in the training step. DDP launches its gradient AllReduce operations inside the backward pass, where they are overlapped with computation and therefore appear as backward-time inflation rather than as a separate reconciliation phase. CPDP, by contrast, still performs synchronized backward steps at reconciliation boundaries and additionally exposes parameter reconciliation after the optimizer step. This helps to explain why CPDP at $K=2$ does not reduce the actual execution time despite the increased accuracy reported with this protocol: the end-to-end time in Table~\ref{tab:cross_site} reflects backward-pass inflation, explicit reconciliation, and residual overhead, not exposed synchronization time alone.

At $K=4$, the periodic methods perform only 12 synchronization events per epoch. Since the per-event cost remains approximately constant, exposed synchronization time drops to about 20.5~s, roughly half of DDP's value. This is consistent with the wall-clock results in Table~\ref{tab:cross_site}, where CPDP at $K=4$ preserves its accuracy advantage while reducing average training time below DDP. CPDP and LocalSGD have nearly identical exposed synchronization time at the same $K$, indicating that their main difference lies in how the synchronized information is used for optimization rather than in measured synchronization cost at this profiler granularity. These exposed-synchronization times quantify synchronization-boundary exposure at the granularity of our profiler; they do not isolate the tensor-only SlowMo update separately.}

{

\paragraph{Intra-site exposed synchronization}

Table~\ref{tab:intra_comm_profile} reports the same direct profiling for the intra-site configurations. Exposed synchronization time is much lower than in the WAN setting, remaining in the 3--10~s range per epoch. At each scale, DDP and CPDP at $K=2$ expose similar total synchronization time, confirming that the reduced frequency of CPDP can be offset by a heavier reconciliation boundary on fast intra-site links. The matched compute floors confirm that the measurements were collected under controlled execution conditions.}

\begin{table}[t]
\centering
\caption{{Direct intra-site profiling of exposed
synchronization time (ResNet-50/CIFAR-100), averaged over three seeds.
Exposed/epoch is the per-epoch exposed synchronization time; the compute floor
is the communication-free baseline. DDP synchronizes every step; CPDP uses
$K=2$. Compute floors are matched between methods at each scale, confirming a
controlled measurement.}}
\label{tab:intra_comm_profile}
\small
\setlength{\tabcolsep}{4pt}
\begin{tabular}{lccc}
\toprule
{Scale} & {Method} & {Exposed/epoch} & {Compute floor} \\
\midrule
g4  & DDP ($K=1$)  & 3.2~s  & 6.7~s \\
g4  & CPDP ($K=2$) & 4.2~s  & 6.6~s \\
\hline
g8  & DDP ($K=1$)  & 9.8~s  & 7.6~s \\
g8  & CPDP ($K=2$) & 9.9~s  & 7.6~s \\
\hline
g12 & DDP ($K=1$)  & 9.6~s  & 8.6~s \\
g12 & CPDP ($K=2$) & 10.0~s & 8.6~s \\
\hline
g16 & DDP ($K=1$)  & 7.4~s  & 9.0~s \\
g16 & CPDP ($K=2$) & 7.7~s  & 9.0~s 
\\
\bottomrule
\end{tabular}
\end{table}

\subsection{Intra-Site Strong Scaling}\label{subsec:intra_scaling}

Table~\ref{tab:scaling} reports accuracy across all three methods 
from 4 to 16 GPUs on the intra-site graffiti cluster. All methods 
use the same linear learning-rate scaling rule (Section~\ref{sec:experimental_setup}).

\begin{table}[htbp]
\centering
\caption{Intra-site strong scaling. ResNet-50/CIFAR-100, 100~epochs. {Mean $\pm$ Std of peak test accuracy over 3 seeds.}}
\label{tab:scaling}
\small
\setlength{\tabcolsep}{2.5pt}
\begin{tabular}{cccc|c}
\toprule

{GPUs} & {DDP} & {LocalSGD} 
& {CPDP} & {CPDP--LocalSGD} \\
\midrule
4 & 79.03~\ensuremath{\pm}~0.28 & 78.70~\ensuremath{\pm}~0.22 
& 79.37~\ensuremath{\pm}~0.14 & \ensuremath{+}0.67 \\
6 & 78.74~\ensuremath{\pm}~0.09 & 78.12~\ensuremath{\pm}~0.34 
& 78.60~\ensuremath{\pm}~0.41 & \ensuremath{+}0.48 \\
8 & 78.94~\ensuremath{\pm}~0.50 & 77.36~\ensuremath{\pm}~0.19 
& 77.97~\ensuremath{\pm}~0.44 & \ensuremath{+}0.61 \\
12 & 78.30~\ensuremath{\pm}~1.25 & 77.12~\ensuremath{\pm}~2.92 
& 78.62~\ensuremath{\pm}~0.64 & \ensuremath{+}1.50 \\
16 & 76.94~\ensuremath{\pm}~0.76 & 66.25~\ensuremath{\pm}~2.06 
& 72.20~\ensuremath{\pm}~1.77 & \ensuremath{+}5.95 
\\
\bottomrule
\end{tabular}
\end{table}

The scaling results show three main trends.
First, CPDP achieves higher mean peak test accuracy than LocalSGD at every evaluated scale, with margins ranging from \ensuremath{+}0.48 percentage points (g6) to \ensuremath{+}5.95 percentage points (g16). This supports an empirical accuracy advantage for the combined CPDP configuration over LocalSGD under the evaluated settings; it does not separately identify the effects of gradient aggregation and SlowMo. The advantage grows with scale: at g16, LocalSGD collapses to 66.25\% under the linear learning rate scaling ($\eta_0=0.4$), while CPDP retains 72.20\%.

Second, CPDP remains competitive with DDP at scales where inter-node communication contributes meaningfully to iteration time. At g4 (single node), CPDP achieves 79.37\% versus DDP's 79.03\%. At g12 (3 nodes), CPDP achieves 78.62\% versus DDP's 78.30\%, with lower inter-run standard deviation (${\pm}0.64\%$ versus ${\pm}1.25\%$). These accuracy statistics describe the evaluated configurations and do not, on their own, allow us to assess the quantitative effect of communications.

Third, the accuracy gap between periodic methods and DDP widens at g16, where the aggressive learning rate ($\eta_0=0.4$) combined with reduced synchronization frequency leads to larger divergence. This is consistent with known effects of large-batch training with linear learning-rate scaling~\cite{lin_dont_2020}. Tuning 
the learning rate scaling strategy at very large global batch sizes remains an open question for all periodic synchronization methods.

All periodic methods in Table~\ref{tab:scaling} use $K=2$. As shown in  Section~\ref{subsec:k_sweep}, higher synchronization periods improve throughput while trading off accuracy at this scale. Furthermore, the g16 degradation observed under linear learning-rate scaling is addressed in  Section~\ref{subsec:lr_sensitivity}, where a more conservative learning-rate setting brings all three methods to
within 0.31 percentage points of each other, with CPDP achieving the highest mean accuracy.

To quantify degradation with scale, Table~\ref{tab:accuracy_drop} reports the accuracy drop relative to the g4 baseline for each method. This allows us to capture variations in model quality when the number of GPUs, the overall batch size, and the scaled base learning rate change simultaneously.

\begin{table}[htbp]
\centering
\caption{Accuracy drop from g4 baseline 
(ResNet-50/CIFAR-100, linear learning-rate scaling). Negative 
values indicate degradation. 
Section~\ref{subsec:lr_sensitivity} shows that the 
g16 degradation for periodic methods is largely 
eliminated with appropriate learning rate tuning.}
\label{tab:accuracy_drop}
\small
\begin{tabular}{lcccc}
\toprule

{Method} & {g4$\to$g6} & {g4$\to$g8} 
& {g4$\to$g12} & {g4$\to$g16} \\
\midrule
DDP & \ensuremath{-}0.29 & \ensuremath{-}0.09 & \ensuremath{-}0.73 & \ensuremath{-}2.09 \\
LocalSGD & \ensuremath{-}0.58 & \ensuremath{-}1.34 & \ensuremath{-}1.58 & \ensuremath{-}12.45 \\
CPDP & \ensuremath{-}0.77 & \ensuremath{-}1.40 & \ensuremath{-}0.75 & \ensuremath{-}7.17 
\\
\bottomrule
\end{tabular}
\end{table}

At small-to-medium scales, all three methods degrade similarly. They diverge at larger scales: at g12, CPDP (\ensuremath{-}0.75~pp) matches DDP (\ensuremath{-}0.73~pp) and improves over LocalSGD (\ensuremath{-}1.58~pp). At g16, CPDP (\ensuremath{-}7.17~pp) is more robust than LocalSGD (\ensuremath{-}12.45~pp), although DDP degrades least overall (\ensuremath{-}2.09~pp), consistent with its fully synchronous optimization.

\begin{figure}[htbp]
\centering
\includegraphics[width=\linewidth]{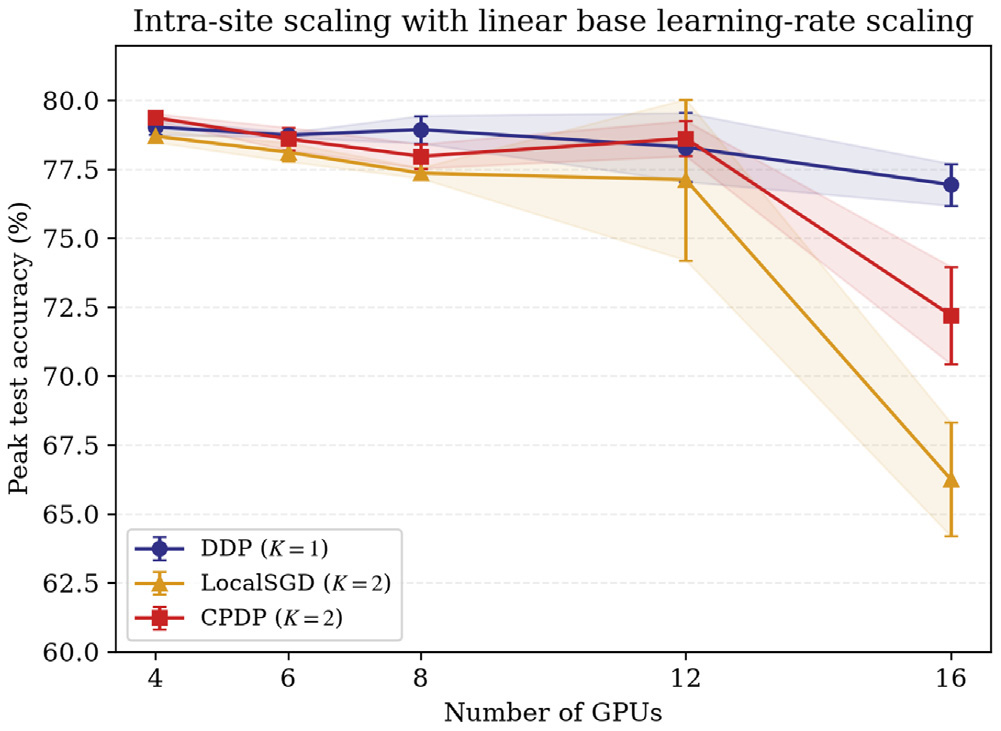}
\caption{Intra-site strong scaling under linear learning-rate scaling
on ResNet-50/CIFAR-100. Accuracy reports mean peak test accuracy
across 3 seeds over the 100-epoch run. DDP maintains the highest
accuracy at larger scales, while CPDP achieves higher mean peak test
accuracy than LocalSGD at each evaluated scale.}
\label{fig:intrasite_scaling}
\end{figure}

Figure~\ref{fig:intrasite_scaling} summarizes these scaling results.
DDP remains the most stable approach under linear learning-rate scaling,
particularly at 16 GPUs.

\subsection{Learning-Rate and Stabilization Sensitivity at Scale}\label{subsec:lr_sensitivity}

The intra-site scaling results in Table~\ref{tab:scaling} show accuracy
degradation at 16 GPUs for periodic methods under aggressive linear
learning-rate scaling ($\eta_0=0.4$). To determine whether this degradation is an
inherent limitation of periodic synchronization or can be recovered under a more
conservative setting, we reduce the learning rate to $\eta_0=0.2$ for all three
methods. For CPDP, we use the main $\beta=0.3$ setting and additionally report a
diagnostic run at $\eta_0=0.2$, $\beta=0.5$ to illustrate the sensitivity to the
SlowMo coefficient.

\begin{table}[htbp]
\centering
\caption{Learning-rate and SlowMo stabilization sensitivity at 16~GPUs.
ResNet-50/CIFAR-100, 100~epochs. Mean $\pm$ standard deviation of peak test
accuracy over 3 seeds. CPDP uses the main $\beta=0.3$ setting in the first and
second columns; the last column reports an additional CPDP diagnostic run at
$\eta_0=0.2$, $\beta=0.5$. ``--'' denotes not applicable.}
\label{tab:g16_tuned}
\small
\setlength{\tabcolsep}{4pt}
\begin{tabular}{lccc}
\toprule

{Method} &
\makecell{Aggressive\\$\eta_0=0.4$} &
\makecell{Rate reduced\\$\eta_0=0.2$} &
\makecell{CPDP diagnostic\\$\eta_0=0.2$, $\beta=0.5$} \\
\midrule
DDP      & 76.94~\ensuremath{\pm}~0.76 & 77.95~\ensuremath{\pm}~0.12 & -- \\
LocalSGD & 66.25~\ensuremath{\pm}~2.06 & 77.86~\ensuremath{\pm}~0.21 & -- \\
\textbf{CPDP} & 72.20~\ensuremath{\pm}~1.77 & \textbf{78.17~\ensuremath{\pm}~0.21} & 74.31~\ensuremath{\pm}~0.18 
\\
\bottomrule
\end{tabular}
\end{table}

\begin{figure}[htbp]
\centering
\includegraphics[width=\linewidth]{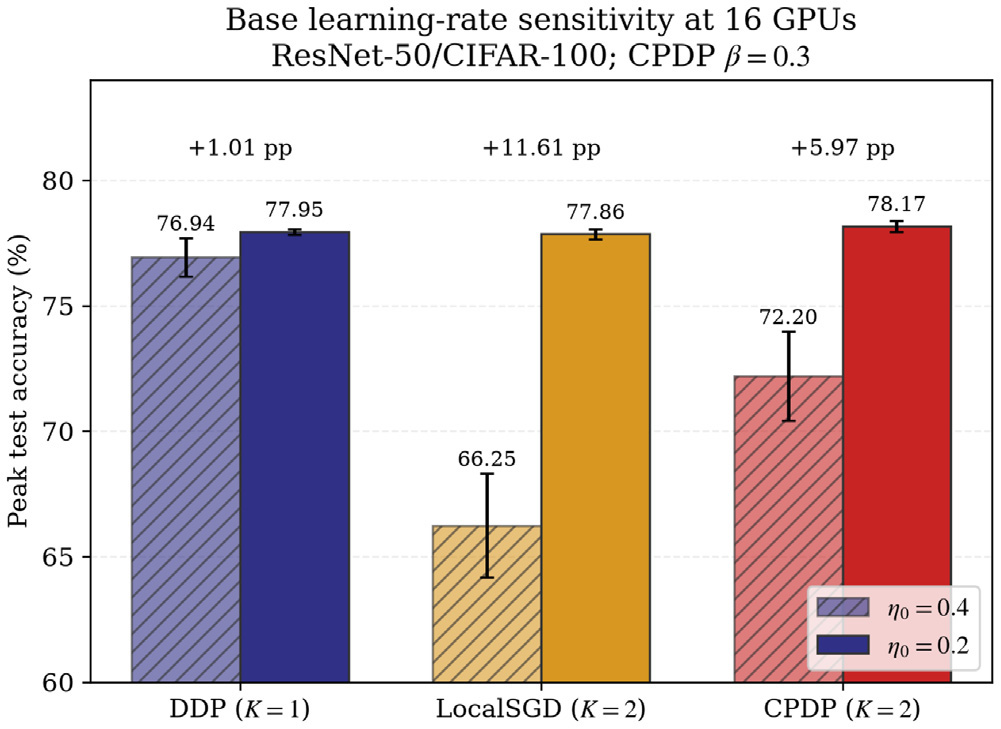}
\caption{Effect of a more conservative learning rate on 16-GPU
ResNet-50/CIFAR-100 training. All methods use their main configuration
(CPDP at $\beta=0.3$); reducing the learning rate from
$\eta_0=0.4$ to $\eta_0=0.2$
improves every method. Table~\ref{tab:g16_tuned} additionally isolates the
SlowMo coefficient via a CPDP diagnostic at $\beta=0.5$.}
\label{fig:lr_sensitivity}
\end{figure}

{
Reducing the learning rate from $\eta_0=0.4$ to $\eta_0=0.2$ substantially improves all
methods at 16 GPUs (see Figure~\ref{fig:lr_sensitivity}). DDP improves from $76.94\pm0.76\%$ to $77.95\pm0.12\%$,
while LocalSGD recovers from $66.25\pm2.06\%$ to $77.86\pm0.21\%$. CPDP also
recovers strongly under the main $\beta=0.3$ setting, improving from
$72.20\pm1.77\%$ to $78.17\pm0.21\%$. This shows that the degradation observed
under aggressive linear learning-rate scaling is not an inherent failure of
periodic synchronization alone; it reflects the coupling between learning rate,
scale, synchronization frequency, and stabilization.

The additional CPDP diagnostic run at $\eta_0=0.2$, $\beta=0.5$ reaches only
$74.31\pm0.18\%$, confirming that the SlowMo coefficient also matters at large
scale. Thus, for CPDP, the large-scale recovery from 72.20\% to 78.17\% is obtained
by reducing the learning rate while keeping the main $\beta=0.3$ setting fixed.
The $\beta=0.5$ diagnostic therefore supports the choice of $\beta=0.3$ in the
main experiments}.

\subsection{Architecture and Dataset Generality}\label{subsec:generality}

To confirm that the observed patterns are not specific to ResNet-50 on CIFAR-100, we evaluate on ViT-S and TinyImageNet.

\paragraph{{{ViT}}-{S} on {{CIFAR}}-100}

Table~\ref{tab:vit} presents results for ViT-S trained with AdamW for 200 epochs.

\begin{table}[htbp]
\centering
\caption{ViT-S/CIFAR-100, 200 epochs. Mean $\pm$ Std of peak test accuracy over 3   seeds.}
\label{tab:vit}
\small
\begin{tabular}{cccc}
\toprule
{GPUs} & {DDP} & {LocalSGD} & {CPDP} \\
\midrule
4 & 60.83~\ensuremath{\pm}~0.22 & 60.40~\ensuremath{\pm}~0.36 & 60.78~\ensuremath{\pm}~0.17 \\
6 & 60.53~\ensuremath{\pm}~0.45 & 59.66~\ensuremath{\pm}~0.60 & 60.04~\ensuremath{\pm}~0.38 \\
8 & 59.51~\ensuremath{\pm}~0.09 & 59.26~\ensuremath{\pm}~0.33 & \textbf{59.90~\ensuremath{\pm}~0.44}
\\
\bottomrule
\end{tabular}
\end{table}

Across scales the three methods remain close, with CPDP at least matching LocalSGD and tracking DDP. At g4 and g6, DDP is marginally highest, with CPDP within 0.05--0.49 percentage points of DDP and above LocalSGD. At g8, CPDP
attains the highest mean accuracy (59.90\% versus 59.51\% for DDP and 59.26\% for LocalSGD), although the difference from DDP is within the inter-run standard deviation. CPDP exceeds LocalSGD at every scale. These results indicate that CPDP remains competitive on AdamW-trained transformer workloads, suggesting that the observed behavior is not restricted to SGD-trained ResNet architectures. 
{For the ViT-S experiments, CPDP uses the same parameter-level reconciliation mechanism as in the ResNet-50 experiments; AdamW moment estimates are not explicitly synchronized.}

\paragraph{{{TinyImageNet}}}

Table~\ref{tab:tinyimagenet} presents results on TinyImageNet (200 classes, {64~\ensuremath{\times}~64} resolution) using ResNet-50.

\begin{table}[htbp]
\centering
\caption{ResNet-50/TinyImageNet, 100~epochs. Mean $\pm$ Std of peak test accuracy over 3 seeds.}
\label{tab:tinyimagenet}
\small
\begin{tabular}{cccc}
\toprule
{GPUs} & {DDP} & {LocalSGD} & {CPDP} \\
\midrule
4 & 68.15~\ensuremath{\pm}~0.07 & 68.27~\ensuremath{\pm}~0.19 & 68.50~\ensuremath{\pm}~0.31 \\
12 & 66.36~\ensuremath{\pm}~0.09 & 64.14~\ensuremath{\pm}~0.49 & 65.83~\ensuremath{\pm}~0.26 
\\
\bottomrule
\end{tabular}
\end{table}

With 4 GPUs (on a single node), the three methods have similar performance within 0.5 percentage points; no loss of accuracy is observed for CPDP with this configuration. At 12 GPUs (3 nodes), all methods degrade relative to the g4 baseline, but by different amounts: DDP drops by 1.79 points, CPDP by 2.67 points, and LocalSGD by 4.13 points. CPDP therefore tracks DDP within roughly half a point at this scale while improving over LocalSGD by \ensuremath{+}1.69 points (65.83\% versus
64.14\%), the largest CPDP--LocalSGD margin among the generality workloads.
Under the evaluated scaling protocol, CPDP preserves accuracy better than LocalSGD on this dataset. This comparison does not isolate inter-node communication as the cause of the accuracy changes.

\begin{table}[htbp]
\centering
\caption{Synchronization period ablation. ResNet-50/CIFAR-100, 8~GPUs, $\eta_0=0.2$, global batch size 512, seed 42. All rows were measured under one launcher on the same two nodes with identical configuration. $K\geq 16$ is an exploratory regime included to map the accuracy boundary of periodic synchronization, not a recommended operating point.}
\label{tab:k_sweep}
\small
\begin{tabular}{lcccc}
\toprule

{Method} & {$K$} 
& {Acc.\ (\%)} & {Time (s)} 
& {Speedup} \\
\midrule
DDP & 1 & 78.53 & 1759 & $1.00\times$ \\
\hline
CPDP & 2 & 78.22 & 1993 & $0.88\times$ \\
CPDP & 4 & 77.68 & 1660 & $1.06\times$ \\
CPDP & 8 & 76.09 & 1496 & $1.18\times$ \\
CPDP & 16 & 69.39 & 1418 & $1.24\times$ \\
CPDP & 32 & 63.70 & 1377 & $1.28\times$ \\
CPDP & 64 & 56.82 & 1351 & $1.30\times$ \\
\hline
LocalSGD & 2 & 77.01 & 1978 & $0.89\times$ \\
LocalSGD & 4 & 76.32 & 1656 & $1.06\times$ \\
LocalSGD & 8 & 76.03 & 1493 & $1.18\times$ \\
LocalSGD & 16 & 71.48 & 1411 & $1.25\times$ \\
LocalSGD & 32 & 63.17 & 1371 & $1.28\times$ \\
LocalSGD & 64 & 59.00 & 1348 & $1.31\times$ 
\\
\bottomrule
\end{tabular}
\end{table}

\subsection{Synchronization Period Ablation}\label{subsec:k_sweep}

\begin{figure}[htbp]
\centering
\includegraphics[width=\linewidth]{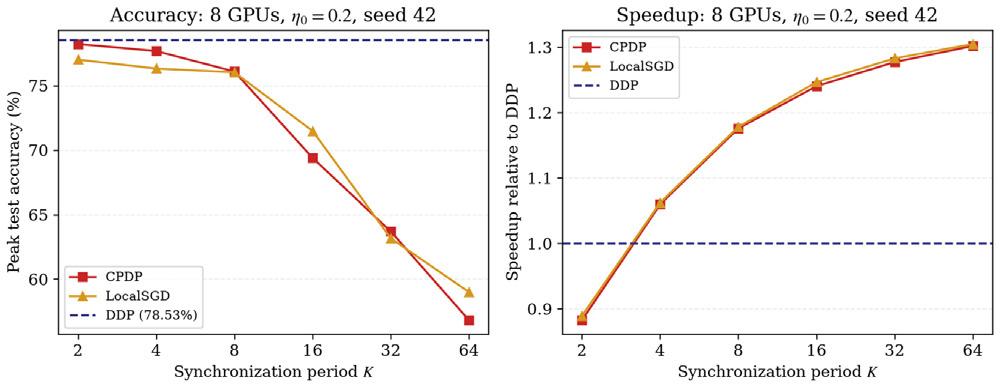}
\caption{Effect of synchronization period $K$ on training accuracy
and speedup relative to DDP (ResNet-50/CIFAR-100, 8~GPUs,
$\eta_0=0.2$, seed 42). Increasing $K$ reduces synchronization
frequency and improves throughput but increases inter-worker drift;
the best accuracy--time trade-off is environment-dependent.}
\label{fig:k_sweep}
\end{figure}

To characterize the trade-off between synchronization frequency, accuracy, and training time, we evaluate CPDP and LocalSGD at multiple synchronization periods on 8~GPUs (Table~\ref{tab:k_sweep}; see Figure~\ref{fig:k_sweep}). This sweep uses seed 42 and is intended to illustrate the trend across synchronization periods, not to provide a statistically validated ranking; {single-seed values may therefore differ slightly from the three-seed means reported in Table~\ref{tab:scaling}.} {All methods in this sweep use the same global batch size (512) and the same $\eta_0=0.2$ schedule.}

The $K$-sweep shows a clear accuracy--time trade-off. In the practical
small-$K$ range, CPDP exceeds LocalSGD by \ensuremath{+}1.21 percentage points at
$K=2$ and \ensuremath{+}1.36 points at $K=4$, while the two methods are nearly
identical at $K=8$. At $K=4$, CPDP reaches 77.68\%, within
0.85 points of DDP, with a $1.06\times$ speedup; at $K=8$, it reaches
76.09\% with a $1.18\times$ speedup. For $K\geq16$, accuracy degrades
sharply while speedup saturates near $1.30\times$, confirming that these
large periods are exploratory rather than practical operating points.
For this workload, the useful range is therefore $K\in\{2,4\}$, with
$K=8$ representing an upper boundary before accuracy loss becomes
substantial.

In intra-site configurations, where communication latency is
substantially lower, $K=4$ or $K=8$ provide stronger
accuracy--throughput trade-offs. A broader WAN-specific sweep beyond $K=4$, potentially combined with adaptive learning-rate schedules or parameter-drift mitigation mechanisms, remains an interesting direction for future work.

\begin{table}[htbp]
\centering
\caption{{SlowMo coefficient sensitivity for CPDP on
ResNet-50/CIFAR-100 at intra-site g4 ($K=2$, $\eta_0=0.1$, 100 epochs).
Values report mean $\pm$ standard deviation over three seeds. Late-epoch std is
computed over the final 10 epochs and averaged across seeds.}}
\label{tab:beta_ablation}
\small
\begin{tabular}{cccc}
\toprule
$\beta$ & {Peak acc.\ (\%)} & {Final acc.\ (\%)} & {Late-epoch std} \\
\midrule
0.0 & 78.74~\ensuremath{\pm}~0.17 & 78.64~\ensuremath{\pm}~0.14 & 0.071 \\
0.3 & 79.38~\ensuremath{\pm}~0.13 & 79.19~\ensuremath{\pm}~0.13 & 0.069 \\
0.5 & 78.90~\ensuremath{\pm}~0.33 & 78.84~\ensuremath{\pm}~0.33 & 0.087 \\
0.7 & 76.77~\ensuremath{\pm}~0.24 & 76.71~\ensuremath{\pm}~0.17 & 0.229 \\
0.9 & 31.56~\ensuremath{\pm}~1.08 & 1.00~\ensuremath{\pm}~0.00  & -- 
\\
\bottomrule
\end{tabular}
\end{table}

{

\paragraph{{{SlowMo}} Coefficient Sensitivity}

Table~\ref{tab:beta_ablation} shows that CPDP is stable for moderate SlowMo coefficients. The best mean peak accuracy is obtained at $\beta=0.3$ (79.38\%), with the lowest late-epoch standard deviation. The difference between $\beta=0.3$ and $\beta=0.5$ is modest and not statistically significant over three seeds (Welch's two-sided $t$-test, $p=0.12$), so we do not claim optimality. However, larger values degrade stability: $\beta=0.7$ reduces accuracy, and $\beta=0.9$ collapses training. We therefore use $\beta=0.3$ as the fixed CPDP configuration in the main experiments and leave a joint search over $\beta$, $K$, and the learning-rate schedule to future work.}

\subsection{Discussion}\label{subsec:discussion}

The experiments show that synchronization frequency is a practical systems parameter for distributed training. Its useful range depends on the network, the workload compute-to-communication ratio, and the amount of parameter drift tolerated between synchronization points. In fast intra-site settings, DDP remains highly competitive because communication is largely overlapped with backpropagation. In this regime, CPDP is most useful when $K$ is large enough to reduce synchronization overhead but still small enough to avoid excessive drift.

The cross-site WAN results show a different behavior. Under the fixed $\eta_0=0.1$ WAN protocol, CPDP at $K=2$ improves accuracy but remains slower than DDP because reconciliation is still frequent. At $K=4$, profiling shows lower exposed synchronization time, and the experiment reports both higher accuracy and lower wall-clock time than DDP. The timing evidence supports the runtime benefit but does not establish a causal effect of WAN pressure on accuracy. This suggests that the best synchronization period is environment-dependent rather than fixed a priori.

\paragraph{Practical guidance for selecting $K$}

Table~\ref{tab:cross_comm_profile} suggests a practical rule: increase
$K$ while the reduction in synchronization frequency outweighs the
higher per-boundary cost, and stop before accuracy degrades. In our
measurements, this points to $K=4$, with $K=2$ as a conservative
starting point.

Across periodic methods, CPDP generally attains higher peak test accuracy than LocalSGD in the practical range $K\leq 4$ under the evaluated settings. Both methods restore parameter equality at reconciliation boundaries; the accuracy results do not demonstrate stronger consistency for CPDP, optimizer-state equality, or a WAN-induced improvement. However, the large-$K$ sweep shows that communication reduction cannot be pushed indefinitely: beyond moderate $K$, local drift dominates and accuracy decreases sharply. CPDP therefore improves the accuracy--time trade-off only when the synchronization period, learning rate, and stabilization coefficient are jointly suitable.

\paragraph{Limitations}

The current study is empirical and limited to 4--16 GPUs, one cross-site WAN pair, and vision workloads. CPDP builds on LocalSGD and SlowMo under the smoothness and bounded-variance assumptions stated in  Section~\ref{subsec:problem_setting}, but we do not provide a new convergence theorem for the combined update under WAN delay, heterogeneous communication cost, and local optimizer states. We also do not compare with gradient compression methods such as PowerSGD~\cite{vogels2019powersgd}, which are orthogonal to synchronization-frequency control. The implementation reconciles model parameters but not optimizer internal states; SGD momentum buffers and AdamW moment estimates remain local between synchronization boundaries. A controlled intervention on the network with appropriate optimization parameters would be necessary to isolate the effect of WAN latency on accuracy. Worker scheduling and resource provisioning~\cite{li_fault_tolerant_2023,zheng_bid_cloud_2015} are complementary future directions; their integration with CPDP has not been evaluated. Future work should also study adaptive $K$ selection, including formulations in which $K$ is treated as a learnable or differentiable control parameter, following broader differentiable-programming and implicit-differentiation approaches~\cite{tao_learning_2026,blondel_elements_2025}, as well as larger NLP workloads, higher-latency WAN settings, and optimizer-state reconciliation.

Synchronization management could also be expressed through AI-specific domain-specific languages (DSLs)~\cite{10.1007/978-3-031-21388-5_4}, inspired by operational DSL frameworks for human--AI workflows~\cite{jiao2026hepscriptdualusedslhumanai} or high-level language and code-generation approaches for parallel scientific computing~\cite{barthou2012qiral}. Exploring such abstractions for synchronization control is a relevant direction for future work.

\section{Conclusion}\label{sec:conclusion}

This paper studied synchronization frequency as a controllable systems parameter for communication-constrained data-parallel training. We evaluated CPDP, a PyTorch-DDP-compatible strategy that alternates local updates with a dual reconciliation step combining gradient AllReduce and SlowMo parameter averaging.

On Grid'5000, CPDP generally achieved higher peak test accuracy
than LocalSGD in the practical small-$K$ range and provided a favorable
accuracy--time operating point in the cross-site WAN setting. Under the fixed $\eta_0=0.1$ WAN protocol, CPDP improved over DDP by \ensuremath{+}2.28 percentage points at $K=2$ but with additional wall-clock time. At $K=4$, it improved over DDP by \ensuremath{+}2.44 percentage points while reducing average training time by 13.8\%. Direct profiling confirmed lower exposed synchronization time at the larger period in this WAN setting. These measurements show a timing advantage; the improvement in accuracy remains conditioned by the learning-rate schedule, the synchronization period, and the SlowMo coefficient.

The results also show that synchronization control is not universally beneficial without tuning. Large synchronization periods and aggressive learning rates can degrade periodic methods, while more conservative learning-rate settings can recover performance. Overall, CPDP demonstrates that synchronization frequency, when combined with drift-aware reconciliation, can improve distributed training behavior under communication constraints.

\section*{Acknowledgements}

This research was supported by the National Center for Scientific and Technical Research (CNRST), Morocco, under the PhD-Associate Scholarship (PASS) program.

Experiments presented in this paper were carried out using the Grid'5000 testbed, supported by a scientific interest group hosted by Inria and including CNRS, RENATER, and several universities and other organizations (see \url{https://www.grid5000.fr}).

\section*{CRediT authorship contribution statement}
\textbf{Imane Ettifouri:} Writing -- original draft, Visualization, Software, Methodology, Investigation, Formal analysis, Data curation, Conceptualization.
\textbf{Mostapha Zbakh:} Writing -- review \& editing, Validation, Supervision, Resources.
\textbf{Claude Tadonki:} Writing -- review \& editing, Validation, Supervision, Resources, Formal analysis.

\section*{Declaration of competing interest}
The authors declare that they have no known competing financial interests or personal relationships that could have appeared to influence the work reported in this paper.

Given their role as Editor, Claude Tadonki had no involvement in the peer-review of this article and has no access to information regarding its peer-review. Full responsibility for the editorial process for this article was delegated to another journal editor.

\section*{Declaration of Generative AI}
Generative AI tools were used only for language editing. All scientific content was developed and verified by the authors.

\section*{Data availability}
The source code and experimental configurations used in this study are publicly available at \url{https://github.com/EmyEtti/CPDP-Distributed-Training}.

\bibliography{refs}

\end{document}